\documentstyle[12pt,a4]{article}
\newcommand{\be}{\begin{equation}}
\newcommand{\ee}{\end{equation}}
\newcommand{\ba}{\begin{eqnarray}}
\newcommand{\ea}{\end{eqnarray}}
\def\1#1{{1 \over {#1}}}
\def\2#1{{{#1} \over 2}}
\def\3#1{{{#1} \over 3}}
\def\4#1{{{#1} \over 4}}
\def\l{\ell}
\def\ie{{\it i}.{\it e}.}

\def\Tr{{\rm Tr}\,}
\def\L{{\rm L}}
\def\R{{\rm R}}
\def\widebar#1{\vbox{\ialign{##\crcr%
          \hskip 1.0pt\hrulefill\hskip 1.0pt%
          \crcr\noalign{\kern-1pt\vskip0.07cm\nointerlineskip}%
          $\hfil\displaystyle{#1}\hfil$\crcr}}}
\def\abs#1{\left| #1 \right|}
\def\wb#1{{\widebar #1}}
\def\swb#1{{\widebar {\scriptstyle #1}}} 
\def\wt#1{{\widetilde #1}}
\def\ul#1{\underline{#1}}
\def\ket#1{\left| #1 \right>}
\def\anoma{anomalous $U(1)$\ }
\def\anom{anomalous $U(1)$}
\def\dGS{\delta_{\rm GS}}
\def\model#1#2{{#1}\,\,;\,{#2}} 
\def\NP#1#2#3{%
{\it Nucl}.~{\it Phys}. {{\bf #1}} (19{#3}) {#2}}
\def\PL#1#2#3{%
{\it Phys}.~{\it Lett}. {{\bf #1}} (19{#3}) {#2}}
\def\PR#1#2#3{%
{\it Phys}.~{\it Rev}. {{\bf #1}} (19{#3}) {#2}}
\def\PRL#1#2#3{%
{\it Phys}.~{\it Rev}.~{\it Lett}. {{\bf #1}} (19{#3}) {#2}}
\def\PTP#1#2#3{%
{\it Prog}.~{\it Theor}.~{\it Phys}. {{\bf #1}} (19{#3}) {#2}}

\def\IJMP#1#2#3{%
{\it Int}.~{\it J}.~{\it Mod}.~{\it Phys}. {{\bf #1}} (19{#3}) {#2}}
\def\PTPS#1#2#3{%
{\it Prog}.~{\it Theor}.~{\it Phys}.~{\it Supplement} %
{{\bf #1}} (19{#3}) {#2}}
\begin{document}

\thispagestyle{empty}
\begin{titlepage}
\begin{flushright}

        {\normalsize
INS-Rep-1179\\
NIIG-DP-96-3\\
December, 1996   }\\
\end{flushright}

\vspace{2cm}

\begin{center}
\Large\bf
``Anomalous'' $U(1)$ Symmetry in Orbifold String Models
\end{center}

\vskip 2cm

\begin{center}
{\large
Tatsuo {\sc Kobayashi}\footnote{
E-mail address: 
kobayast@ins.u-tokyo.ac.jp}
 and Hiroaki {\sc Nakano}$^{\dagger}$\footnote{%
E-mail address: nakano@muse.hep.sc.niigata-u.ac.jp}
}
\\
\vspace{0.5cm}
{\it
Institute for Nuclear Study, University of Tokyo, \\
Midori-cho, Tanashi, Tokyo 188, Japan}
\\
\vspace{0.5cm}
{$^{\dagger}$%
\it Department of Physics, Niigata University, \\
Niigata 950-21, Japan}
\end{center}

\vskip 2cm

\begin{abstract}
``Anomalous'' $U(1)$ gauge symmetry with Green-Schwarz anomaly
cancellation mechanism is discussed in the orbifold construction
of four-dimensional heterotic string models.
Some conditions are given as criteria
to have ``anomalous'' $U(1)$ in orbifold string models.
In particular, ``anomalous'' $U(1)$ is absent
if the massless twisted matter has no mixing
between visible and hidden sectors
or if a certain type of discrete symmetries are found.
We then give a general procedure
for classifying orbifold models with ``anomalous'' $U(1)$
and for identifying the ``anomalous'' $U(1)$ basis.
We illustrate our procedure in $Z_3$ and $Z_4$ orbifold models.
According to our procedure,
the classification of ``anomalous'' $U(1)$ can be reduced to
the classification in the absence of a Wilson line.
We also discuss discrete symmetries left unbroken
after the ``anomalous'' $U(1)$ breaking.
This includes a possible relation
between ``anomalous'' $U(1)$ and discrete $R$-symmetries.
\end{abstract}
\vfill
\end{titlepage}
\setcounter{footnote}{0}

\section{Introduction}

Superstring theory is a promising candidate of the unified theory.
Lacking for the dynamical principle
determining the true string vacuum,
many efforts have been devoted to construct
semi-realistic string models directly in four dimensions.
The physical content turns out to be rich enough that
there have been found many semi-realistic models.
It is important to extract such model-independent properties
characteristic to string models that are not shared
by usual field theoretic approach to unified theories.

The original surprise in superstring theory was the anomaly cancellation
found in ten dimensions by Green and Schwarz\cite{GS}.
As the result of the global consistency condition,
modular invariance, of the world-sheet theory,
we have many examples of four-dimensional string models
which show a miraculous pattern of gauge anomaly cancellation.
In particular, we often have string models
possessing so-called ``anomalous'' $U(1)$ gauge symmetry.
This $U(1)$ gauge symmetry has non-vanishing contributions
to anomalies from chiral fermions.
In fact, these anomalies are canceled
via four-dimensional counterpart of Green-Schwarz mechanism, \ie,
by assigning the nonlinear transformation to the axion-like field.
Since such anomaly cancellation mechanism of \anoma
is intimately related to the consistency of string theory,
we can expect characteristic and interesting consequences 
from this mechanism.
[A discussion related to this point can be found
in Ref.~\cite{moduli}, where an \anoma is used
to derive a constraint on non-perturbative effects in string theory.]

Many string models are known which possess the \anoma gauge symmetry.
These, however, have been analyzed in a model-dependent way
and no criteria for the appearance of
the \anoma gauge symmetry has been completely clarified yet.
Which class of string models does give rise to the \anom?
At present,
we need a detailed analysis of massless spectrum in each model
and struggle with $U(1)$ charges
to see whether a string model contains an \anom.
This situation is quite unsatisfactory
not only for theoretical purpose
but also for phenomenological applications that we mention shortly.
Therefore a systematic analysis is desirable.
To do this is the main motivation of the present work.

The \anoma gauge symmetry
with Green-Schwarz anomaly cancellation mechanism
is very interesting by itself
and has actually received renewed interests.
The presence of mixed $U(1)$-gravitational anomaly in particular
implies that its $U(1)$ generator $Q$ is not traceless $\Tr Q\neq 0$
leading to the generation of Fayet-Iliopoulos term\cite{FI}.
This in turn breaks\footnote{
Otherwise, the supersymmetry itself breaks down.
This possibility is potentially interesting if
we consider the ``dual'' theory in which the supersymmetry breaking
would show up in a non-perturbative manner.
} the \anoma automatically\cite{DSW}.
The breaking scale is calculated\cite{DSWii} to be
just below the string scale:
\be
M_{{\rm A}}^2 = \frac{\Tr Q}{192\pi^2}M_{{\rm st}}^2 \ .
\ee
Unlike the conventional scheme of grand unifications,
we need not entangle with complicated Higgs structure.
Many interesting applications of \anoma come
from this automatic breaking.

There are several contexts in which
applications of the \anoma are addressed.
The \anoma has most widely been discussed in constructing
a string model with a realistic gauge group and matter content;
the \anoma breaking sometimes triggers
further breaking of gauge symmetries
and therefore provides a way of reducing the rank of gauge group%
\cite{degenerate,reduce,U1 charge}.
It was also pointed out\cite{IB} that
we have an interesting way of calculating the weak mixing angle
without any grand unification symmetries
since the anomaly cancellation condition relates the normalization
of gauge couplings to anomaly coefficients.
Also as several authors observed,
the breaking scale $M_{{\rm A}}$ is quite impressive for
the origin of various hierarchical structures in particle physics.
Utilizing the ratio $M_{{\rm A}}/M_{{\rm st}}$,
higher dimensional couplings could explain
hierarchical structures in the fermion masses and mixing angles.
Actually there are several proposals\cite{%
texture:non-anomalous,texture:IR,texture:others}
for realistic fermion mass matrices
based on a family $U(1)$ symmetry, which is often anomalous,
while stringy selection rules were used in Ref.~\cite{texture2}.
{}Furthermore,
generalizing $U(1)$ $D$-term contributions
in supergravity models\cite{Di,KMY},
it was argued\cite{anoma,KK,mass:others}
that the \anoma gives a new source to
non-universality of soft breaking scalar masses
and the cosmological constant through Fayet-Iliopoulos $D$-term,
both of which should be taken into account when we are to have
the universal scalar mass and vanishing cosmological constant.

A remarkable possibility has recently been pointed out
on the basis of these developments:
An \anoma can be used to construct
a model with supersymmetry breaking%
\cite{SUSY-breaking:mech,SUSY-breaking:model}.
Also in cosmological context,
some authors have argued\cite{cosmology,inflation}
that it can play a role in constructing a model of inflation.
A possible solution to doublet-triplet splitting problem
has also been suggested\cite{splitting}.

Now,
it is desirable to take a systematic approach 
to string models with \anoma
in order to further explore these possibilities.
The first issue to be discussed is to clarify the condition
under which we have an \anoma gauge symmetry in string models.
Then the second issue is the pattern of the \anoma breaking:
to characterize the flat directions
along which the \anoma breaking occurs
and to see what kind of consequences we have after such breaking.

In this paper, we shall mainly discuss the first issue
by examining orbifold string models.
We find that
the appearance of \anoma is constrained by several reasons.
We show in particular that an orbifold model possesses an \anoma
only if it contains massless twisted matter
which leads to the mixing between the visible and hidden sectors.
We also give several examples of discrete symmetries
that forbid the appearance of \anom.
Moreover,
we argue that
the analysis of orbifold models in a Wilson line background
can essentially be reduced to the analysis in the absence of
a Wilson line.
We then give a general procedure for classifying
orbifold string models which possess \anom.

Concerning the second issue,
we give a brief discussion on discrete symmetries
unbroken after the \anoma breaking.
These discrete symmetries would be relevant to
phenomenological problems
such as suppression of dangerous couplings.
We also suggest a possible relation of the \anoma
to discrete $R$ symmetries.

Our interest on the discrete $R$ symmetry originates
not only from phenomenological but also string theoretical aspects.
It was argued\cite{DS,naturalness} that
a certain discrete $R$ symmetry can protect (0,2) string vacua
against the instability\cite{DSWW}
due to the world-sheet instanton effects.
In particular,
Dine and Seiberg constructed an example of such $R$ symmetry
in a (2,2) string vacuum (the standard embedding of $Z_3$ orbifold)
which guarantees the existence of the flat directions
along which the (2,2) vacuum can be deformed into (0,2) ones.
Other aspects concerning the stability of (0,2) vacua
are recently discussed in Refs.~\cite{%
Distler-Greene,Distler-Kachru,Silverstein-Witten}.
This issue is beyond the scope of this paper.

This paper is organized as follows.
In the next section,
we give a review of the orbifold construction.
In particular we recall the expressions for
the mass formula and generalized GSO projection operator
which determine massless spectrum.
We also recall the formulas for the $U(1)$ level and charges.
In section three,
after recapitulating the universal nature of $U(1)$ anomalies
in string theories,
we discuss the conditions for a $U(1)$ to be anomalous.
Our procedure for classifying and identifying the \anoma
is presented in section four.
We illustrate it by working out $Z_3$ orbifold models.
[A classification of $Z_4$ orbifold models is given in Appendix~B.]
In section five, we briefly discuss the issues
on the flat directions and \anoma breaking.
The final section is devoted to concluding remarks.
Some concrete examples for the orbifold models
with the \anoma are given in appendices.

\section{Orbifold Construction}

In this section, we give a review of the orbifold construction
of four-dimensional string models\cite{orbifold:original}.
{}For details, see Refs.~\cite{Orbi1,Orbi2,Orbi3}.

The Hilbert space of heterotic string theory is
a tensor product of the right-moving sector,
which is responsible for space-time supersymmetry, and
the left-moving sector, which gives rise to gauge symmetries.
The right moving sector is a superconformal field theory
of the $(4+6)$-dimensional string coordinates $X^{\mu=0,1,2,3}$ and
$X^{i=1,2,3}$ (and their complex conjugates $X^{\swb{i}}$)
as well as their world-sheet superpartners, NSR fermions
$(\psi^i,\psi^{\swb{i}};\psi^{\mu})$.
The latter are conveniently bosonized, $\psi^t=-ie^{iH^t}$, into $H^t$
($t=1,2,3$ and $t=4,5$ correspond to
six- and four-dimensional parts, respectively)
which are related to Cartan part of $SO(9,1)$ current algebra.
The momenta $p^t$ of $H^t$ span an $SO(9,1)$ weight lattice
$\Gamma_{SO(9,1)}$.
Neveu-Schwarz sector (space-time boson) corresponds to
vectorial weights $p^t_v$ with integer entries
and Ramond sector (space-time fermion) to
spinorial weights $p^t_s$ with half-integer entries.

The conformal field theory of the left moving sector includes
four-dimensional string coordinates $\wb{X}^{\mu}$,
and six-dimensional parts $\wb{X}^i$ and $\wb{X}^{\swb{i}}$.
Adopting the bosonic formulation for the gauge sector,
we also have anti-chiral coordinates parametrizing
the maximal torus of $E_8\times E_8'$,
$\wb{X}^{\hat{I}}=(\wb{X}^I,\,\wb{X}^{I'})$,
whose momenta $P^{\hat{I}}=(P^I,\,P^{I'})$
span an $E_8\times E_8'$ root lattice $\Gamma_{E_8\times E_8'}$
($I,I'=1,\cdots,8$).
The root vectors of $E_8$ satisfy $(P^I)^2=2$ and are represented as
\ba
P^I&=&\left(\underline{\pm 1, \pm 1, 0,0,0,0,0,0}\right)
\equiv\left(\underline{\pm 1, \pm 1, 0^6}\right)\ ,
\label{vector root}
\\
P^I&=&\left(\pm\1{2}, \pm\1{2}, \pm\1{2}, \pm\1{2},
        \pm\1{2}, \pm\1{2}, \pm\1{2}, \pm\1{2}\right)
\label{spinor root}
\ea
with the even number of minus signs for the latter case.
Here and hereafter the underline means permutations
and repeated entries are indicated by superscripts.

In a toroidal compactification to four dimensions,
the resultant space-time supersymmetry charges
form four-dimensional representation of
$SU(4)\simeq SO(6)$ subgroup of $SO(9,1)$.
In the orbifold construction,
we divide the six-dimensional torus $T^6$
by its discrete isometry, called point group $P$,
in order to have precisely $N=1$ supersymmetry.
We choose the $SO(9,1)$ weight $p_Q^t$ of the surviving supercharge
to be $p_Q^t=\pm (1,1,1\,;\,\pm 1,\pm 1)/2$ (with even $-$'s).

We concentrate on symmetric $Z_N$ orbifolds.
Then the six-dimensional part of the string coordinates
combines into $X^i(z,\wb{z})=X^i(z)+\wb{X}^i(\wb{z})$
which become the coordinates of a torus $T^6={\bf R}^6/\Lambda$
through dividing by a root lattice $\Lambda$
of a semi-simple Lie algebra.
The orbifold modding is realized through a further division of 
this torus by the point group $P=Z_N$, which is 
generated by a twist $\theta$, $\theta^N=1$.
These operations can equivalently be expressed through division
by a space group $S=(\Lambda,P)$; $T^6/P\simeq {\bf R}^6/S$.

In order to achieve the modular invariance,
we should include twisted sectors.
In $Z_N$ orbifold models,
we have the $k$-th twisted sector $T_k$ ($k=1,2,\cdots, N-1$)
as well as the untwisted sector $U$.
In the $T_k$-sector, the six-dimensional coordinates satisfy
\be
X^i(e^{2\pi i}z,e^{-2\pi i}\wb{z})=e^{2\pi ikv_i}X^i(z,\wb{z})+e^i \ ,
\label{boundary condition:X}
\ee
where
$e^{2\pi iv_i}$ is the eigenvalue of the twist $\theta$ on $X^i$
and the vector $e^i$ belongs to
the defining lattice $\Lambda$ of the torus $T^6$.
The right-moving world-sheet supersymmetry then requires
\be
H^i(e^{2\pi i}z)=H^i(z)+2\pi k v^i \ .
\label{boundary condition:H}
\ee
The shift vector $v^t=(v^i;\,0,0)$ in $\Gamma_{SO(9,1)}$ should
be orthogonal to the weight $p^t_Q$ of the surviving supercharge.
{}For the Abelian embedding of the twist into gauge group,
we associate a shift $V^{\hat{I}}$ to the rotation $\theta$
as well as a Wilson line $a^{\hat{I}}$ to the translation $e^i$
of the space group. Then the boundary condition of
sixteen-dimensional gauge coordinates is
\be
\wb{X}^{\hat{I}}(e^{-2\pi i}\wb{z})=
\wb{X}^{\hat{I}}(\wb{z})+2\pi kV^{\hat{I}}
+2\pi m^ia_i^{\hat{I}} \ ,
\label{boundary condition:gauge}
\ee
where the integer $m^i$ labels fixed points.
The modular invariance restricts the possible choice of
shift vectors in $\Gamma_{E_8\times E_8'}$
and Wilson lines according to
\be
N\left[
\sum_{i=1}^3\left(v^i\right)^2-
\sum_{{\hat I}=1}^{16}\left(V^{\hat I}+m^ia_i^{\hat I}\right)^2\right]
=~\hbox{ even integer } \ .
\label{modular: shift}
\ee
All the possible shifts 
$v^t=(v^i\,;\,0,0)$ and $V^{\hat{I}}=(V^I\,;\,V^{I'})$
are known for each $Z_N$ orbifold construction\cite{shift}.
The simplest choice is the standard embedding;
$V^{\hat{I}}=(v^i,0^5\,;\,0^8)$, $a^{\hat{I}}=0$.
We also refer to the following type of the shift
as {\it a quasi-standard embedding}
\be
V^{\hat{I}}=(v^i,0^5\,;\,V^{I'}) \ .
\label{quasi}
\ee

On-shell string states are created by vertex operators
acting on the vacua of (super)conformal field theories
on the string world sheet,
$V_{\R}(z)\ket{0}_{\R}\otimes V_{\L}(\wb{z})\ket{0}_{\L}$.
The internal part of the vertex operators takes the form
\be
V_{\R}\sim e^{ip_{\R}^iX^i}e^{ip^tH^t} \ , \qquad
V_{\L}\sim e^{ip_{\L}^i\swb{X}^i}e^{iP^{\hat I}\swb{X}^{\hat I}} \ .
\ee
In twisted sectors, we should drop the momenta $p_{\R,\L}^i$ 
and replace the $SO(9,1)$ and $E_8\times E_8$ momenta
with shifted ones defined by
\be
\wt{p}^t\equiv p^t+kv^t \ , \qquad
\wt{P}^{\hat{I}}\equiv P^{\hat{I}}+kV^{\hat{I}}+m^ia_i^{\hat{I}} \ .
\label{shifted mom}
\ee

Each twisted sector $T_k$ has several subsectors
corresponding to fixed points labeled by
the twist $\theta^k$ and $e^i$
(by the conjugacy class of the space group, to be precise).
The vertex operator for such twisted state includes
the twist field \cite{orbifold-CFT}, $\sigma(k,e)$, 
which creates the twisted vacuum $\ket{k,e}=\sigma(k,e)\ket{0}$
and expresses the twisted boundary condition
(\ref{boundary condition:X})
of the internal string coordinates $X^{i=1,2,3}$.
These twist fields contribute to the conformal dimension of
the ground state in the $k$-th twisted sector by an amount
\be
c^{(k)}\equiv\1{2}\sum_{i=1}^3
        \eta_{(k)}^i\left(1-\eta_{(k)}^i\right) \ , \qquad
\eta^i_{(k)}\equiv \abs{kv^i}-{\rm Int}\,\abs{kv^i} \ .
\label{casimir}
\ee
We also recall that
there arise some complications concerning the twisted vacua
for higher twisted sectors ($k=2,\cdots,N-2$)
in non-prime order orbifolds ($N=4,6,8,12$),
which are relevant for the generalized GSO projection.
In this case,
the fixed points of the higher twist $\theta^k$ do not necessarily
fixed by the single twist $\theta$ but transform into each other.
Therefore we have to take their linear combination
to form an eigenstate under the single twist\cite{KO:linear comb}.
In Ref.~\cite{Orbi3},
such eigenstates were explicitly constructed
with their eigenvalues $e^{i\gamma}$ under the single twist.

Mass formulas and physical states are most easily
described in light-cone gauge\rlap.\footnote{
In our convention,
we simply remove the last component of the $SO(9,1)$ momentum
to get the transverse $SO(8)$ momentum.
}
Mass formulas are obtained by counting the conformal dimensions
of vertex operators and are given, for the right- and left-moving
$T_k$-sectors (untwisted sector $U$ for $k=0$), respectively, by
\ba
\1{8}m^2_{{\rm R}}
&=&\1{2}\sum_{i=1}^3\left(p^i_{\R}\right)^2
 + \1{2}\sum_{t=1}^4\left(\wt{p}^t\right)^2
 + N_{\R}^{(k)}-\1{2}+c^{(k)} \ ,
\label{mass:right}
\\
\1{8}m^2_{{\rm L}}
&=&\1{2}\sum_{i=1}^3\left(p^i_{\L}\right)^2
 + \1{2}\sum_{\hat{I}=1}^{16}\left(\wt{P}^{\hat{I}}\right)^2
 + N_{\L}^{(k)}-1+c^{(k)} \ ,
\label{mass:left}
\ea
where oscillator numbers $N_{\R,\L}^{(k)}$ take
the fractional value which is a multiple of $1/N$.
Alternatively we can express the contribution from sixteen-dimensional
gauge part in terms of the Kac-Moody algebra in the following way.
Let $\prod_aG_a$ be the gauge group.
If the state with the (shifted) momentum $\wt{P}^{\hat{I}}$
transforms in the representation $\otimes_aR_a$,
then we have a formula
\be
\1{2}\sum_{\hat{I}=1}^{16}\left(\wt{P}^{\hat{I}}\right)^2
= \sum_ah_a\left(R_a\right)
\equiv \sum_a{C_2(R_a) \over k_a + C_2(G_a)} \ ,
\label{h}
\ee
where $k_a$ is a Kac-Moody level and $C_2(R_a)$ ($C_2(G_a)$) is 
a Casimir for the $R_a$ (adjoint) representation of the group $G_a$.
If the group $G_a$ is Abelian,
the conformal dimension of the state carrying its $U(1)$ charge $Q_a$
is given by $h_a(Q_a)=Q^2_a/k_a$.

The physical states of orbifold models should be invariant
under the full action of the orbifold twist.
{}For the untwisted sector, this leads to modulo integer conditions
\be
P^{\hat{I}}V^{\hat{I}}-p^tv^t=P^{\hat{I}}a^{\hat{I}}= 0 \ ,
\label{utmass}
\ee
from which
we can identify gauge groups and massless untwisted matter contents.
Massless gauge bosons correspond to
the $SO(8)$ weights $p_v^t=(0^3\,;\,\pm 1)$
and so satisfy $P^{\hat{I}}V^{\hat{I}}=P^{\hat{I}}a^{\hat{I}}=0$
while massless untwisted matter fields to $p_v^t=(\ul{1,0^2}\,;\,0)$.
{}For instance, models with the quasi-standard embedding
(\ref{quasi}) always contain an $E_6$ gauge group.
Generically, a $U(1)$ factor gauge group appears
corresponding to the non-vanishing element
of the shift or Wilson lines
although some combination of non-vanishing elements may
correspond to the Cartan part of a non-Abelian group.
When a $U(1)$ corresponds to a basis vector $V_Q^{\hat I}$,
the level $k_Q$ and the charge $Q$
of the state with a momentum $\wt{P}^{\hat I}$ are given,
respectively, by\cite{realistic,U1 charge}
\ba
k_Q&=&2\sum_{\hat I=1}^{16}\left(V_Q^{\hat I}\right)^2 \ ,
\label{level:def}
\\
Q&=&\sum_{{\hat I}=1}^{16}V_Q^{\hat I}\wt{P}^{\hat I} \ .
\label{charge:def}
\ea

The physical states in twisted sectors are singled out
by the generalized GSO projection operator\cite{Orbi1,GSO,Orbi3}
\be
G_k\equiv\1{N}\sum_{h=0}^{N-1}\left(\Delta_k\right)^h
\nonumber
\ee
so that only the states with $\Delta_k=1$ survive.
Here we recall from Ref.~\cite{Orbi3} the expression
for the operator $\Delta_k$
in the absence of a Wilson line:\footnote{
The expression in the presence of Wilson lines
can be found in Refs.~\cite{KO:WL,Orbi3}.
}
\ba
\Delta_k&\equiv&
e^{i\gamma}e^{2\pi i\left(N_{\R}+N_{\L}\right)}e^{2\pi i\Theta_k} \ ,
\label{GSO:delta}
\\
\Theta_k&\equiv&
\2{k}\left[\sum_i\left(v^i\right)^2
-\sum_{\hat{I}}\left(V^{\hat{I}}\right)^2\right]
+\left[\sum_{\hat{I}}\wt{P}^{\hat{I}}V^{\hat{I}}
-\sum_i\wt{p}^iv^i\right] \ ,
\label{GSO:theta}
\ea
where the first term in $\Theta_k$,
which is important for non-standard embedding cases, expresses
$Z_N$-transformation property of twisted ground states
while the second expresses that of vertex operators.
The phase $e^{i\gamma}$ described above should be kept in the case
of the higher twisted sectors of non-prime order orbifolds.
{}For the case of $Z_{N=3,7}$ orbifold models,
the GSO projection is trivial and 
the level matching condition $m_{\R}^2=m_{\L}^2$ is known to be enough
to guarantee the modular invariance.

\section{Constraints on ``Anomalous'' $U(1)$}

String models without left-moving world-sheet supersymmetry,
\ie, (0,2) models, often lead to \anoma symmetry.
[Illustrative examples are given in appendix~A.]
We wish to have a criterion for the appearance of \anom.
In this section,
we examine the massless conditions described in the previous section
and find that the appearance of \anoma is constrained
by several reasons.

\subsection{Visible-Hidden Sector Mixing in Twisted Sector}

The first category of such constraints comes from
the universal nature of Green-Schwarz mechanism.
Suppose that we have a gauge group $U(1)_A\times\prod_aG_a$
and the $U(1)_A$ is anomalous.
In the presence of \anom,
the K\"ahler potential of the dilaton-axion chiral multiplet $S$
is given at one-string loop\footnote{
This K\"ahler potential includes the term which becomes
the Fayet-Iliopoulos term after the dilaton develops the VEV.
Notice that
the nonlinear transformation (\ref{dilaton:trf}) explains why
we can have the Fayet-Ilopoulos term in supergravity Lagrangian
{\it without} $U(1)$ $R$ symmetry
unlike the conventional wisdom in supergravity\cite{FI:R}.
We also note that
any other one-string loop effects are not considered here
including the possible kinetic mixing
between several $U(1)$'s\cite{U1 mixing}.
} by\cite{DSW}
\be
K=-\ln\left(S+S^{\dagger}-\dGS\,V_A\right) \ ,
\label{dilaton kahler}
\ee
where $V_A$ is the vector multiplet of $U(1)_A$
and $\dGS$ is a constant related
to the mixed gravitational anomaly;
it is $\Tr Q_A=96\pi^2\sqrt{k_A}\dGS$
where $Q_A$ and $k_A$ are the charge and level of $U(1)_A$,
respectively.
On the other hand, the gauge kinetic function $f_a$ of
a factor group $G_a$ is given at string tree level by $f_a=k_aS$,
\be
{\cal L}_{{\rm gauge}}=
\1{4}\sum_ak_a\int d^2\theta\, S W^{\alpha (a)}W_\alpha^{(a)}
+ \hbox{H.c.}\ ,
\label{gauge kinetic}
\ee
where the summation is taken over all gauge groups
including the anomalous $U(1)_A$.
The existence of the axion-like coupling of ${\rm Im}\,S$
in eq.~(\ref{gauge kinetic}) enables us to cancel
the pure $U(1)_A^3$ and mixed $U(1)_A$ - $G_a^2$ anomalies
as well as the mixed gravitational one
by combining the nonlinear transformation of
the dilaton-axion field with the $U(1)_A$ gauge transformation
\ba
V_A&\longrightarrow&V_A+\2{i}\left(\Lambda-\Lambda^\dagger\right) \ ,
\label{vector:trf}
\\
S&\longrightarrow&S+\2{i}\dGS\,\Lambda \ ,
\label{dilaton:trf}
\ea
where $\Lambda$ is a parameter chiral superfield.
In any modular invariant string theory in four-dimensions,
all the $U(1)$ anomalies should be canceled
in this manner\cite{SW:anomaly}.
Hence their anomaly coefficients should satisfy
the following universality relation:
\be
\1{k_a}\mathop{\Tr}_{G_a}T(R)Q_A
=\1{3}\Tr Q_A^3 =\1{24}\Tr Q_A\,\left(\,\equiv 8\pi^2\dGS\,\right) \ ,
\label{universal:without U1}
\ee
where $2\,T(R)$ is the index of the representation $R$,
and we have rescaled $Q_A$ so that $k_A=1$.
We refer to this relation as the universal GS relation.

It is important to realize that
when the gauge group contains several $U(1)$'s,
each $U(1)$ should satisfy the universal GS relation.
This property,
which can explicitly be confirmed by examples given in Appendices,
follows from the uniqueness of the \anom;
we can always find a unique $U(1)_A$ which may be anomalous
so that other $U(1)$'s are anomaly free\cite{SW:anomaly,U1 charge}.
Then the universal GS relation for this $U(1)_A$ reads
\be
\1{k_a}\mathop{\Tr}_{~G_a}T(R)Q_A
=\Tr Q_B^2Q_A
=\1{3}\Tr Q_A^3 =\1{24}\Tr Q_A=8\pi^2\dGS \ ,
\label{universal}
\ee
where $Q_B$ is the charge of any non-anomalous $U(1)$
and has been rescaled so that $U(1)_B$ has level one.
By using this equation,
we can show that any linear combinations,
$U(1)_{\alpha}=\alpha U(1)_A+\beta U(1)_B$ and
$U(1)_{\beta}=\beta U(1)_A-\alpha U(1)_B$ with $\alpha^2+\beta^2=1$,
satisfy the same relations as eq.~(\ref{universal})
with rescaled GS constants, $\alpha\dGS$ and $\beta\dGS$,
respectively. 
We thus see that 
any $U(1)$ satisfies the universal GS relation
even if it does not coincide with the true \anom.
This fact is useful in the following discussion.

We can derive several constraints on \anoma
from the universal nature of anomaly in string theory.
The basic observation is that if a $U(1)$ symmetry has
no mixed $U(1)$ - $G_a^2$ anomaly for a certain group $G_a$,
then all the $U(1)$ - $G_b^2$ anomalies should vanish for any $G_b$.
Therefore we can judge whether a $U(1)$ is anomalous or not
by examining just a single type of anomaly.
Practically it is easiest to examine the mixed $U(1)$ anomaly
with the largest gauge group $G_{\l}$
since the massless conditions tightly restrict the appearance
of nontrivial representations of $G_{\l}$.
{}For instance,
massless matter fields in nontrivial representations of
$E_8$ are forbidden and anomalies involving $E_8$ always vanish.
This explains why all (2,2) models which lead to an unbroken $E_8$
can not have an \anoma symmetry.

To derive further constraints,
we write a gauge group in the form
\be
G=G_{{\rm vis}}\times G_{{\rm hid}}
 = \left[\prod_a G_a \times U(1)^m \right] \times
   \left[\prod_b G'_b\times U(1)'^n\right] \ ,
\label{vis-hid}
\ee 
where the visible- and hidden-sector groups
$G_{{\rm vis}}$ and $G_{{\rm hid}}$
are originated from $E_8$ and $E_8'$, respectively.
Generally it is possible that some massless states transform
nontrivially under both of visible and hidden groups.
If the model contains no such state, we call it the model with
the complete separation of the visible and hidden sectors.
Then the universal nature (\ref{universal}) clearly tells us that
models with the complete separation have no \anoma symmetry.
Actually we have even stronger constraints:
A $U(1)_a$ gauge group in the visible sector is anomaly free
if there is a hidden group $G_b'$ so that
any massless $G_b'$-charged field has vanishing $U(1)_a$ charge.
Notice, furthermore, that
there is no mixing between visible and hidden sectors
in the untwisted sector:
all massless untwisted fields in the visible sector
have vanishing $E_8'$-momentum $P^{I'}=0$
and are neutral under $G_{{\rm hid}}$ and vice versa.
The mixing can arises only through
the shift (\ref{shifted mom}) of $E_8\times E_8'$ momenta.
Hence we can restrict ourselves to twisted sectors
and conclude that
{\it if the visible-hidden sector mixing is absent
for massless twisted matter fields, the model has no \anoma symmetry}.

We now apply the above constraint to show that many models
whose gauge group contains an $E_7$ or $E_6$ do not have an \anom.
{}First consider models with an $E_7$.
Here we restrict ourselves to the models with $k_a=1$.
The expression~(\ref{h}) tells us that
the massless condition (\ref{mass:left}) forbids
the representation with the conformal dimension larger than $1$.
Then, other than the singlet, massless matter fields can only
belong to the representation $\ul{56}$,
which has the conformal dimension $h(\ul{56})=3/4$
as is seen from $C_2(E_7)=18$ and $C_2(\ul{56})=57/4$.
Actually there are many models in which
massless $\ul{56}$'s are forbidden in twisted sectors
or have vanishing $U(1)$ charges.
An example is the $T_1$-sector of $Z_3$ orbifold models
in which $c^{(1)}=1/3$ and thus the field in $\ul{56}$
can not satisfy the massless condition (\ref{mass:left}).
Another example is the $T_2$-sector of $Z_4$ orbifold models
in which $c^{(2)}=1/4$.
Even if a massless $\ul{56}$ appears in this sector,
it can not have a non-vanishing charge for any $U(1)$ group 
since the massless condition is already saturated,
$h(\ul{56})+c^{(2)}-1=0$.
The same is true in any twisted sectors of
$Z_3$, $Z_4$, $Z_6$-I, $Z_7$ and $Z_8$-I orbifold models
in the classification of Refs.~\cite{Orbi2,Orbi3}.
Therefore, if the hidden gauge group contains an $E_7$,
all $U(1)$ symmetries in the visible sector are anomaly free
for the $Z_N$ orbifold models
other than $Z_6$-II, $Z_8$-II, $Z_{12}$-I and -II.

Note, however, that since our constraint here utilizes
the absence of the visible-hidden sector mixing, 
it does not exclude the \anoma in the same sector as an $E_7$.
The possible hidden gauge groups which include an $E_7$ is
$E'_7\times U(1)'$ other than anomaly free $E'_7\times SU(2)'$.
The $U(1)'$ accompanied by the $E'_7$ is anomalous
if the untwisted sector\footnote{
In the untwisted sector with $P^{\hat I}V^{\hat I}\equiv 1/2$
which is $N=2$ subsector,
the degeneracy factor is important.
} contains a massless $\ul{56}$.
Actually an explicit analysis shows that
it {\it is} always anomalous before Wilson lines are turned on.

A similar analysis can be extended to
the models with an $E_6$ gauge group.
The $E_6$ group with $k=1$ allows only the $\ul{27}$
and its conjugate representations since we have 
$C_2(E_6)=12$, $C_2(\ul{27})=26/3$ and $h(\ul{27})=2/3$.
Therefore the same result as in the models with an $E_7$ can be derived
for the $Z_3$ orbifold models containing an $E_6$ gauge group.
On the other hand, other $Z_N$ orbifold models are less constrained
even if they contain an $E_6$ gauge group
(although the appearance of massless $\ul{27}$'s is not so often
since their conformal dimension is large).
Here we only make some comments on the models
with the quasi-standard embeddings (\ref{quasi}).
Even if we restrict ourselves to this class of models,
there are some examples in which an \anoma appears
even in the opposite sector to the $E_6$.
If we restrict further to the cases without a Wilson line,
however, an explicit analysis shows that
the $E_6$ group derived from the quasi-standard embeddings
does not have any nontrivial twisted matter fields and
all $U(1)$'s in the opposite sector to the $E_6$ are anomaly free.
Especially in the case of $Z_7$ orbifold models,
all the quasi-standard embeddings do not lead to the \anoma
in the visible as well as hidden sectors.

One may wonder whether
the above constraints were not so powerful
since the realistic model building involves the Wilson lines
which do not allow such a large gauge group as an $E_7$ or $E_6$.
As we shall see in the next section, however,
the origin of \anoma can be traced back to the cases
without a Wilson line
and so the constraints given here turn out to be already powerful.

\subsection{Discrete Symmetries}

The second category of constraints comes from
discrete symmetries of a spectrum.
In this section, we examine the mass formulas
and describe several examples of such symmetries
that forbid the appearance of an \anom.
As is clear from the discussion on $U(1)$ basis 
given around eq.~(\ref{universal}),
we can examine the visible and hidden sectors separately
and so we concentrate on the visible-sector gauge group
coming from the first $E_8$.

As noted above,
a $U(1)$ gauge symmetry generically corresponds to
the non-vanishing element of the shift $V^I$ or Wilson lines $a^I$;
such non-vanishing element breaks the $E_8$
and generically corresponds to an unbroken $U(1)$ basis.
On the other hand,
a vanishing element of the shift and Wilson line generically
corresponds to the Cartan basis of a non-Abelian gauge group.
Only exception\cite{reduce,U1 charge} is the case with the total shift
of the form $kV^I+m^ia_i^I=(*^7,0)$,
where $*$ indicates non-zero entries.
In any case, if $J$-th and $K$-th components of 
the shift and Wilson lines vanish simultaneously,
they correspond to the Cartan parts of a non-Abelian group.

In this subsection,
we are mainly interested in the $U(1)$'s
that correspond to non-vanishing elements of a Wilson line.
It is instructive, however, to reconsider why
the anomaly cancels for the $U(1)$ which corresponds to 
a vanishing element of the shift and Wilson line.
Suppose that the $J$-th components of
the shift and Wilson line are zero: $V^J=a^J=0$.
Then we observe that $\wt{P}^J=P^J$ and that
{\it the mass formulas (\ref{mass:left}),
physical state conditions (\ref{utmass})
and generalized GSO projection operator
(\ref{GSO:delta}), (\ref{GSO:theta})
are all invariant under the transformation which reverses
the sign of the $J$-th component of $E_8$-momenta}:
\be
P^I=\left(\cdots, P^J, \cdots\,\right)
\ \longrightarrow\ 
\wb{P}^I\equiv\left(\cdots, -P^J, \cdots\,\right) \ .
\label{Z2 symmetry}
\ee
This implies that if the state with $P^I$ is massless and physical,
there exists the massless physical state with $\wb{P}^I$
and thus the $U(1)$ charges corresponding to the $J$-th
Cartan generator sum up to vanish
\be
\sum P^J=0 \ ,
\ee
where the summation is taken over all massless states.
In this way,
the absence of the anomaly for the $U(1)$'s corresponding to
vanishing elements of the shift and Wilson line can be understood
on the basis of the discrete symmetry of the spectrum.

In fact, there is a subtlety in the above discussion.
If the $E_8$-momentum $P^I$ corresponds
to a spinorial root (\ref{spinor root}),
the operation (\ref{Z2 symmetry}) flips the chirality.
Then we need another component $V^K=a^K=0$
so that the simultaneous change
$P^J\rightarrow -P^J$ and $P^K\rightarrow -P^K$
preserves the chirality.
Otherwise, we have to examine whether or not
the states with a spinorial root exist and contribute to the anomaly.

Now let us extend the above argument to show that
there is no anomaly in the $U(1)$
corresponding to a non-vanishing element of Wilson lines
if the shift and Wilson lines are orthogonal to each other.
[This includes the case with a vanishing shift.]
We mainly consider the $Z_3$ orbifold models with
a single independent Wilson line $a^I$ for simplicity.
Because of the orthogonality, we can use
an $E_8$ transformation so that $V^J=0$ for $a^J\neq 0$.

We examine the untwisted and twisted sectors separately.
{}For the untwisted sector,
precisely the same argument as above applies.
If an $E_8$-momentum $P^J$ satisfies
the massless conditions (\ref{utmass}) for untwisted fields,
the momentum $\wb{P}^I$ whose $J$-th component is replaced
with $-P^J$ also satisfies it.
This is the symmetry from which we conclude that
the charges of the $U(1)$ corresponding to
the non-vanishing element $a^J\neq 0$ of the Wilson line
sums up to vanish in the untwisted sector.
{}For the twisted sector,
the momenta of states are shifted as in eq.~(\ref{shifted mom}).
On each two-dimensional $Z_3$ orbifold,
there are three subsectors corresponding to three fixed points
with $e^i=m\,e^i_{SU(3)}$ ($m=0,\,\pm 1$)
in eq.~(\ref{boundary condition:X}),
where $e^i_{SU(3)}$ is the simple root of the $SU(3)$ lattice.
In the subsector corresponding to $m=0$,
the massless condition (\ref{mass:left}) is symmetric
under $P^{\hat J} \rightarrow -P^{\hat J}$
and the situation is similar to the untwisted sector.
Interestingly, if $P^{\hat J}$ satisfies
the massless condition (\ref{mass:left}) for $m=1$,
$-P^{\hat J}$ satisfies it for $m=-1$
and therefore the contributions to the anomaly cancel
between $m=1$ and $m=-1$ subsectors.

Note that if there exists the massless state with a spinorial root,
we must take care of the chirality as noted before.
We need another component $V^K=0$
so that the simultaneous change
$P^J\rightarrow -P^J$ and $P^K\rightarrow -P^K$ preserves the
chirality.
This is always possible in $Z_3$ orbifold models
and therefore we have $\sum\wt{P}^J=\sum\wt{P}^K=0$.

We thus see that
if the Wilson line is orthogonal to the shift,
there exists a discrete symmetry which guarantees
the cancellation of anomaly for the $U(1)$ basis
that corresponds to non-vanishing elements of the Wilson line.
In this case, the \anoma basis is related
only with non-vanishing elements of the shift.
{\it If the $E_8$ breaking by the shift does not produce an \anom,
there is no anomaly even if several $U(1)$'s appear
by switching on the Wilson lines orthogonal to the shift}.

{}For other orbifold models, we can find similar symmetries.
{}For instance, the two-dimensional $Z_4$ orbifold has
two subsectors corresponding to $m=0,1$.
The corresponding Wilson line $a^I$ satisfies
$2a^I=0$ up to $\Gamma_{E_8}$
as was shown in Refs.~\cite{KO:WL,Orbi3}.
This implies that
the subsector with $m=1$ is equivalent to one with $m=-1$
as far as the massless condition is concerned.
If there is no massless state with a spinorial $\wt{P}^I$,
this equivalence will lead to a symmetry of the massless spectrum.

If a Wilson line is not orthogonal to the shift,
the symmetry described above is broken.
Even in such a case, however,
it happens that a similar symmetry is found
if the Wilson line and shift satisfy a certain relation.
Let us take the $Z_3$ orbifold model
with $V^J=2/3$ and $a^J=2/3$ as an example.
The twisted states in the subsector with $m=-1$ have $\wt{P}^J=P^J$
while the states in the subsectors with $m=0$ and $m=1$ have
the shifted momenta, $\wt{P}^J=P^J+2/3$ and 
$\wt{P}^J=P^J+4/3=(P^J+2)-2/3$, respectively.
Therefore the sum of the quantum number $\wt{P}^{\hat J}$ vanishes
for the twisted states as in the above discussion
(although such a cancellation does not always work
in the untwisted sector).
A similar cancellation between twisted states can be found
in several combinations of the shift and Wilson line.

Next we study another symmetry in the massless spectrum.
Consider the $Z_7$ orbifold model with the standard embedding
$V^{\hat{I}}=\left(1,2,-3,0^5;0^8\right)/7$.
This model has the gauge group
$G=E_6\times U(1)_1\times U(1)_2\times E_8'$
whose $U(1)$ bases can be taken as
\be
V_1^I=\frac{\sqrt{3}}{2}\left(1,-1,0,0^5\right) \ , \qquad
V_2^I=\frac{1}{2}\left(1,1,-2,0^5\right) \ ,
\label{Z7 basis:U}
\ee
or equivalently,
\be
\wt{V}_1^I=\frac{1}{2\sqrt{7}}\left(5,-4,-1,0^5\right) \ , \qquad
\wt{V}_2^I=\frac{\sqrt{3}}{2\sqrt{7}}\left(1,2,-3,0^5\right) \ .
\label{Z7 basis:T}
\ee
Although the absence of $U(1)$ anomaly in this model can be proved
by other reasons such as the existence of the $E_8'$
or (2,2) superconformal symmetry,
this model has remarkable symmetries
that guarantee the anomaly cancellation.

\begin{table}[t]
\begin{center}
\begin{tabular}{@{\vrule width 1pt \ }r|cc|cc@{\ \vrule width 1pt}}
\noalign{\hrule height 1pt}
        &\multicolumn{2}{c|}{$\ul{27}$}
        &\multicolumn{2}{c@{\ \vrule width 1pt}}{$\ul{1}$}\\
\cline{2-5}
        & $Q_1 $ & $Q_2$ & $Q_1/\sqrt{3}$ & $Q_2/\sqrt{3}$ \\
\noalign{\hrule height 1pt}
\hline\hline
\noalign{\hrule height 1pt}
$U_1$ & $\2{\sqrt{3}}$ & $\2{1}$ & $-1$ & $0$\\
\hline
$U_2$ & $-\2{\sqrt{3}}$ & $\2{1}$ & $\2{1}$ & $-\2{\sqrt{3}}$\\
\hline
$U_4$ & $0$ & $-1$ & $\2{1}$ & $\2{\sqrt{3}}$\\
\noalign{\hrule height 1pt}
\end{tabular}
\end{center}
\caption[$U(1)$ charges of untwisted matters]%
{$U(1)$ charges of untwisted matters in $Z_7$ standard embedding}
\label{table:U1 charge:untwisted}
\end{table}

\begin{table}[hbt]
\begin{center}
\begin{tabular}{@{\vrule width 1pt \ }l|cc|cc|cc|cc@{\ \vrule width 1pt}}
\noalign{\hrule height 1pt}
        &\multicolumn{2}{c|}{$\ul{27}$~~($7\times 1$)}
        &\multicolumn{2}{c|}{$\ul{1}$ ~~($7\times 1$)}
        &\multicolumn{2}{c|}{$\ul{1}$ ~~($7\times 2$)}
        &\multicolumn{2}{c@{\ \vrule width 1pt}}{$\ul{1}$~~($7\times 4$)}\\
\cline{2-9}
        & $\sqrt{7}\wt{Q}_1$
        & $\sqrt{7}\wt{Q}_2$
        & $\frac{\sqrt{7}}{2\sqrt{3}}\wt{Q}_1$
        & $\frac{\sqrt{7}}{2\sqrt{3}}\wt{Q}_2$
        & $\frac{\sqrt{7}}{3}\wt{Q}_1$
        & $\frac{\sqrt{7}}{3}\wt{Q}_2$
        & $\frac{\sqrt{7}}{\sqrt{3}}\wt{Q}_1$
        & $\frac{\sqrt{7}}{\sqrt{3}}\wt{Q}_2$ \\
\noalign{\hrule height 1pt}
\hline\hline
\noalign{\hrule height 1pt}
  $T_1$ & $-\2{1}$        & $-\2{\sqrt{3}}$
        & $-\2{\sqrt{3}}$ & $-\2{1}$
        & $\2{1}$         & $-\2{\sqrt{3}}$
        & $0$             & $1$ \\
\hline
  $T_2$ & $-\2{1}$        & $+\2{\sqrt{3}}$
        & $0$             & $1$
        & $-1$            & $0$
        & $+\2{\sqrt{3}}$ & $-\2{1}$ \\
\hline
  $T_4$ & $1$             & $0$
        & $+\2{\sqrt{3}}$ & $-\2{1}$
        & $\2{1}$         & $+\2{\sqrt{3}}$
        & $-\2{\sqrt{3}}$ & $-\2{1}$ \\
\noalign{\hrule height 1pt}
\end{tabular}
\end{center}
\caption[$U(1)$ charges of twisted matters]%
{$U(1)$ charges of twisted matters in $Z_7$ standard embedding}
\label{table:U1 charge:twisted}
\end{table}

The untwisted sector has three subsectors $U_1$, $U_2$ and $U_4$
which are classified by the value $\sum_ip^iv^i=1/7,\,2/7$ and $4/7$.
Each untwisted subsector has a $\ul{27}+\ul{1}$
as the massless matter.
In each twisted sector $T_{1,2,4}$, we have
\be
7\times \left[\ul{27}+\ul{1}+2\times\ul{1}+4\times\ul{1}\right] \ ,
\ee
where the first $\ul{27}$'s come from non-oscillated states,
and remaining singlets from oscillated states
with $N_{\L}=1/7,\,2/7,\,4/7$, respectively.
The $U(1)$ charges of these fields are shown
in Table~\ref{table:U1 charge:untwisted}
for the untwisted states in the basis (\ref{Z7 basis:U})
and in Table~\ref{table:U1 charge:twisted}
for the twisted states in the basis (\ref{Z7 basis:T}).
The states in each column of both tables form a triplet
and thus the anomaly cancels between them.
Actually the massless spectrum of this model is symmetric under
\be
X^1\ \longrightarrow\ 
X^2\ \longrightarrow\ 
X^3\ \longrightarrow\ X^1 \ ,
\label{Z3}
\ee
which respectively rotates the untwisted subsectors $U_{1,2,4}$
as well as the twisted sectors $T_{1,2,4}$
with the same oscillator number into each other.
We can find a similar symmetry in some other orbifold models
(even with non-standard embeddings).
The partial list includes the untwisted states for $Z_8$-I and 
$Z_{12}$-I orbifold models with quasi-standard embeddings,
whose defining shifts are
$v^i=1/8(1,2,-3)$ and $1/12(1,4,-5)$, respectively.

We also note that
the above $Z_7$ orbifold model has another type of symmetry.
We see from Table~\ref{table:U1 charge:twisted} that
the $U(1)$ charges of the singlets with different oscillator numbers
sum up to vanish in a single twisted sector.
[The numbers in parentheses of Table~\ref{table:U1 charge:twisted}
show the multiplicity of the states.]
This is the symmetry within each subsector
corresponding to the fixed point and might be interesting
when we include a Wilson line,
whose role is to resolve the degeneracy between the fixed points.

\section{Classification of ``Anomalous'' $U(1)$ in Orbifold Models}

In this section we examine in detail $Z_3$ orbifold models
and give a procedure for classifying the models with an \anom.
We work out the models in the absence of a Wilson line
and extend the analysis to the case with a Wilson line.
Although we deal only with $Z_3$ orbifold models,
our procedure is general and can be applied to other models.

Let us first recapitulate the classification of
$Z_3$ orbifold models in the absence of a Wilson line
with attention to the visible-hidden sector mixing.
Modular invariant pairs of shifts $(V^I\,;\,V^{I'})$ are
classified into five types including a trivial one as
\ba
& {\rm No.~0}: & (3V^I;\,3V^{I'})= (0^8;0^8) \ , \nonumber \\
& {\rm No.~1}: & (3V^I;\,3V^{I'})= (2,1,1,0^5;0^8)\ , \nonumber \\
& {\rm No.~2}: & (3V^I;\,3V^{I'})= (2,1,1,0^5;2,1,1,0^5) \ ,
\label{shift5} \\
& {\rm No.~3}: & (3V^I;\,3V^{I'})= (1,1,0^6;2,0^7)\ , \nonumber  \\
& {\rm No.~4}: & (3V^I;\,3V^{I'})= (2,1^4,0^3;2,0^7)\ , \nonumber 
\ea
up to $E_8 \times E_8'$ automorphisms\rlap.\footnote{
One can subtract $E_8\times E_8'$ roots
so that the total shift has the length less than one.
See also Ref.~\cite{reduction}.
}
The first model is a trivial one with
unbroken $E_8 \times E_8'$ gauge group and no massless matter field.
The second one (model No.~1) corresponds to the standard embedding
with $E_6\times SU(3)\times E_8'$ gauge group.
The massless twisted matter fields are
\be
(27,1;\,1') \quad {\rm for}\quad N_{\L}=0 \ , \qquad 
( 1,3;\,1') \quad {\rm for}\quad N_{\L}=\3{1} \ .
\label{matter:No.1}
\ee
This model has no visible-hidden sector mixing.
The third one (model No.~2) corresponds to a quasi-standard embedding
and leads to the gauge group $E_6\times SU(3)\times E_6'\times SU(3)'$.
This model has the visible-hidden sector mixing
due to the massless twisted matter
\be
(1,3;\,1',3') \quad {\rm for}\quad  N_{\L}=0 \ ,
\label{matter:No.2}
\ee
but this mixing does not contribute to anomaly.
Thus the models No.~0, 1 and 2 contain no U(1) gauge group
and are anomaly free.
On the other hand, an \anoma does arise in the models No.~3 and 4 
as we describe in detail in Appendix~A.

We now proceed to models in the presence of a Wilson line.
Each subsector corresponding to the fixed point labeled by
the intergers $m^i$ has the total shift of the form
\be
\left(\,V^I+m^ia^I_i\,;\,V^{I'}+m^ia^{I'}_i\,\right) \ .
\label{total shift}
\ee
It is remarkable that
{\it any modular invariant pair of the total shift
is equivalent to one of five shifts} (\ref{shift5})
{\it up to $E_8 \times E_8'$ automorphisms}.
This property enables us to have a simple classification of \anom.
We call the shift $V^{\hat I}=(V^I\,;\,V^{I'})$
which is equivalent to the total shift (\ref{total shift})
of the subsector under consideration as {\it an equivalent shift}.
Similarly we call such a model without a Wilson line
that has the equivalent shift of the subsector under consideration
as {\it an equivalent model}.
All the complication comes from the fact that
such equivalent shifts of subsectors
can be different from each other:
{}For example it is possible that
the subsector labeled by $m^i=(0,0,0)$ has
an equivalent shift of No.~0 type
while the equivalent shift for $m^i=(1,0,0)$ is of No.~1 type.

{}For an illustrative purpose, let us first discuss the model
with the shift No.~1 and a single Wilson line:
\ba
3V^{\hat I}\!
&=&(2,1,1,0,0,0,0^2)(0,0,0,0^5)'
\ , \nonumber \\
3\,a_1^{\hat I}
&=&(0,0,0,2,1,1,0^2)(2,1,1,0^5)' \ .
\label{example1}
\ea
The gauge group $E_6\times SU(3)_1\times E_8'$ of the model No.~1
is broken by this Wilson line to 
\be
SU(3)_1\times SU(3)_2\times SU(3)_3\times SU(2)\times U(1)
\times E_6' \times SU(3)' \ .
\label{gauge:example1}
\ee
The twisted sector of this model has three subsectors
corresponding to $m^1=0,~\pm 1$.
The subsector with $m^1=0$ has the same structure
as the twisted sector of the model No.~1;
the massless matter fields appear in the representation of
$E_6\times SU(3)_1\times E_8'$ as in eq.~(\ref{matter:No.1}).
Actually the $E_6$ is broken to
$SU(3)_2 \times SU(3)_3 \times SU(2) \times U(1)$
under which the $\ul{27}$ decomposes into $(3,3,2)_1+(3,3,1)_{-2}$.
Nevertheless,
it is important to realize that
matter fields appear as if they form an $E_6$ multiplet.
It is then clear that
this subsector does not contribute to anomaly
since the $U(1)$ in eq.~(\ref{gauge:example1}) is a part of the $E_6$
in this subsector.
Similarly,
the subsector with $m^1=1$ has the equivalent shift of No.~2 type
and the massless matter fields appear in the representation of
$E_6\times SU(3)_1\times E_6'\times SU(3)'$.
In this case, all the fields are singlet under the broken $E_6$
and so has vanishing $U(1)$ charges.
Thus this subsector does not contribute to anomaly
in spite of the visible-hidden sector mixing
due to the field (\ref{matter:No.2}).
[Even if the $SU(3)$ is broken to
$SU(2)\times U(1)$ or $U(1)^2$ by another Wilson line,
the field $(1,3\,;\,1',3')$ does not produce any anomalies.]
The subsector with $m^1=-1$ is also of No.~2 type
and has no contribution to anomaly.
Thus this model has only No.~1 and 2 types of
twisted massless matter fields and does not contain the \anom.

An important lesson from the above example is that
{\it the massless matter content in any subsector
is precisely the same as in the corresponding twisted sector
of the equivalent model}.
This can directly be checked by looking at the massless condition.
[Degeneracy factors also coincide with each other
when counted in the corresponding subsector in the equivalent model.]
As far as each subsector is concerned,
the effect of Wilson lines is just to decompose the representation
of matter fields according to the unbroken gauge group,
and the situation is quite similar to
the conventional Higgs mechanism of grand unified theories.
This is not true for a whole model, of course,
since each subsector can correspond to a different type of models.
Note also that some of untwisted matter fields are projected out
by including a Wilson line.

Now it is clear what criteria we have
for the appearance of an \anom:
If the total shift (\ref{total shift}) is equivalent
to one of the shifts No.~0, 1 and 2 listed in eq.~(\ref{shift5})
for all the subsectors $m^i=0,~\pm 1$,
then the model contains no \anom.
On the other hand, a model contains an \anoma
if there is a subsector whose total shift is equivalent
to the shift No.~3 or 4.
If there are several such subsectors,
the \anoma is their linear combination
(as long as the cancellation between them does not occur).

Let us illustrate how our procedure for finding an \anoma works
in concrete examples. First consider the model with 
\ba
3V^{\hat I}\!&=&(2,1,1,0^5)(2,1,1,0^5)' \ ,
\nonumber \\
3\,a_1^{\hat I}&=&(-2,0,0,0^5)(0,-1,-1,0^5)' \ .
\label{example2}
\ea
This model has the gauge group
$$
SO(10)\times SU(2) \times U(1)_1 \times U(1)_2 \times 
SO(10)'\times SU(2)' \times U(1)'_1 \times U(1)'_2 \ ,
$$
whose $U(1)$ bases we take as 
\ba
\hskip -1cm
&&U(1)_1\,:~~Q_1 =(0,1,1,0^5)(0^8)' \ , \qquad 
 U(1)_1'\,:~~Q_1'=(0^8)(0,1,1,0^5)' \ ,
\nonumber \\
\hskip -1cm
&&U(1)_2\,:~~Q_2 =(1,0,0,0^5)(0^8)' \ , \qquad 
 U(1)_2'\,:~~Q_2'=(0^8)(1,0,0,0^5)' \ . 
\label{U1basis}
\ea
The subsectors with $m^1=0$ and $m^1=-1$ have the total shifts
equivalent to No.~2 type shift and do not contribute to anomaly.
The equivalent shift in the subsector with $m^1=1$ is of No.~3 type
and thus an \anoma arises from this subsector.
The massless matter fields appear in the representation
of broken $E_7\times U(1)_1\times SO(14)'\times U(1)_2'$,
and as is shown in Appendix~A, the $U(1)_1$ is anomalous
while the $U(1)_2'$ is anomaly free.
The remaining $U(1)_2$ and $U(1)_1'$ are anomaly free
since they are contained in the $E_7$ or $SO(14)'$.
In this way, one can conclude that this model contains an \anom,
whose basis is given by $(0,1,1,0^5)(0^8)'$.

Next
we discuss the model in which two subsectors contribute to anomaly.
An example of such model is given by
\ba
3V^{\hat I}\!&=&(0,1,1,0^5)(2,0,0,0^5)' \ ,
\nonumber \\
3\,a_1^{\hat I}&=&(-2,-1,-1,0^5)(-2,-1,-1,0^5)' \ .
\label{example3}
\ea
This model has the same gauge group as the previous one
and we take the same $U(1)$ bases (\ref{U1basis}).
The subsector with $m^1=0$ has the equivalent shift of No.~3 type
and the massless matter fields form the multiplets
under the broken $E_7\times U(1)_1\times SO(14)'\times U(1)'_2$.
In particular,
there is the visible-hidden mixing due to $(1;\,14')_{(2,-1)/3}$,
which contributes to anomaly so that $\Tr_{SO(10)'}Q_1\neq 0$.
In addition,
the subsector with $m^1=1$ also has the total shift
equivalent to No.~3 type shift.
The massless matter fields appear in the representations
under the broken $SO(14)\times U(1)_2\times E_7'\times U(1)_1'$
in this subsector.
In particular,
the existence of $(14;\,1')_{(1,-2)/3}$
causes $\Tr_{SO(10)}Q'_1\neq 0$.
The subsector with $m^1=-1$ is of No.~2 type
and does not contribute to anomaly.
As a result, 
the true \anoma is a linear combination of
$U(1)_1$ and $U(1)_1'$, whose coefficients are determined
by calculating $\Tr_{SO(10)}Q'_1$ and $\Tr_{SO(10)'}Q_1$
as $Q_1-Q_1'$.

The third example is the model in which
the anomaly cancels between two subsectors:
\be
3V^{\hat I}=3\,a_1^{\hat I}
=(2,1,1,1,1,0^3)(2,0^7)' \ .
\label{example4}
\ee
The gauge group is $SU(9)\times SO(14)'\times U(1)'$
as in the model No.~4, but the inclusion of the Wilson line
removes all the massless untwisted matter.
The subsector with $m^1=-1$ has the model No.~0 as an equivalent model
and contains no massless matter fields.
The matter content of $m^1=0$ subsector is the same
as in the twisted sector of the model No.~4
and is given by nine copies of $(9,1')_{2/3}$.
In the $m^1=+1$ subsector,
the total shift is $V^{\hat I}+a_1^{\hat I}=2V^{\hat I}$
which is just the shift of the $T_2$-sector in the model No.~4.
Therefore this subsector contains nine copies of $(9^*,1')_{-2/3}$
as is seen from the general fact\cite{CP} that
the matter content in the $T_2$-sector is
$CPT$ conjugate of that in the $T_1$-sector.
These fields cancel the anomaly arising from the $m=0$ subsector.

We can extend these analysis to
more general cases as well as other orbifold models.
Actually such an analysis proceeds as follows:
\begin{enumerate}
\renewcommand{\labelenumi}{\labelenumii}
\item[(i)]
Classify all the models before a Wilson line is included:
In each model,
work out its twisted matter content
and identify the basis of an \anoma and its charges.
\item[(ii)]
Turn on Wilson lines.
{}For each subsector of a twisted sector $T_k$,
\begin{enumerate}
\item
identify the equivalent model\rlap.\footnote{
A careful analysis shows that
the GSO projection in the presence of a Wilson line
results in the same degeneracy factor
as in the corresponding twisted subsector of the equivalent model.
}
This subsector has no contribution to anomaly
if the $T_k$-sector of the equivalent model
does not contain the massless matter 
which contributes to the visible-hidden sector mixing.
\item
``pull back'' the \anoma basis $V_{Q^{(0)}}$ in the equivalent model
to get the basis $V_Q$ of the \anoma in this subsector.
Schematically,
\def\llongrightarrow{-\!\!\!-\!\!\!-\!\!\!-\!\!\!\longrightarrow}
\def\llongleftarrow{\longleftarrow\!\!\!-\!\!\!-\!\!\!-\!\!\!-}
\ba
\hskip -2cm
\hbox{shift in $E_8\times E'_8$ lattice}
:\ \ \! kV^{\hat I}+m^ia_i^{\hat I}\ 
\mathop{{\llongrightarrow}}^{R_{E_8\times E_8'}}&&\!\!\!\!\!\!
kV^{\hat I}
\nonumber\\
&&\!\!\!\Downarrow\ \hbox{step (i)}
\label{scheme}
\\
\hskip -2cm
\hbox{\anoma basis}~
:\ \ \qquad V_Q^{\hat I}\qquad
\mathop{\llongleftarrow}_{R_{E_8\times E_8'}^{-1}}&&\!\!\!\!\!
V_{Q^{(0)}}^{\hat I}
\nonumber
\ea
\item
decompose the matter content of the $T_k$-sector of the equivalent model
according to the unbroken gauge group
in the presence of Wilson lines.
In general, the ``pull back'' operation as in (\ref{scheme})
is necessary also in this step.
\end{enumerate}
\item[(iii)]
Put these subsectors as well as the untwisted sector together
to get the whole model\rlap.\footnote{
It might be interesting to observe that
the construction of an orbifold model in a Wilson line background
resembles the construction of a fiber bundle.
}
The true basis of the \anoma is given by a linear combination of
$U(1)$ bases obtained in the step (b).
The massless matter content and $U(1)$ charges are
obtained from the results of the step (c).
\end{enumerate}
A classification of modular invariant pairs of the shifts
$(V^I\,;\,V^{I'})$ is already available\cite{shift,Orbi2}
and it is straightforward to identify the basis of $U(1)$
which has the visible-hidden sector mixing contributing to anomaly.
As an example, a classification of visible-hidden sector mixing
in $Z_4$ orbifold models is given in Appendix~B,
where several new features are observed.

We finish this section by two remarks.
{}Firstly, the absence of an \anoma can be established
by examining only twisted sectors
in spite of the fact that
the untwisted sector may contribute to anomaly.
This is a special case of more general phenomena:
As in the example (\ref{example3}),
the true \anoma is a linear combination of several $U(1)$'s
when several subsectors contribute to the anomaly.
In such a case,
the true basis of the \anoma can be determined by calculating
the mixed anomaly between the visible and hidden sectors
and therefore by examining only twisted sectors.
These are the consequences of the universal nature of anomaly
in string theory as described in section~3.1.
Secondly,
a cancellation of anomaly may occur
between several subsectors which contribute to the anomaly.
We should remark that
such a cancellation can be understood, in some cases,
by the discrete symmetry described in  section~3.2.
An example is provided by the model (\ref{example4}).
We note also that
the absence of anomaly in the model (\ref{example1})
can be understood by the discrete symmetry,
\ie, by the orthogonality of the Wilson line to the shift.
In this way,
the analysis based on the discrete symmetries plays a role
complimentary to the analysis in this section.

\section{``Anomalous'' $U(1)$ Breaking and Discrete Symmetries}

The \anoma gauge symmetry breaks automatically
once the dilaton VEV is fixed by yet unknown mechanism.
There exists a flat direction along which some scalar fields
develop the VEV's to cancel the Fayet-Iliopoulos term.
Then the next issue to be discussed is
along which flat direction the \anoma breaking occurs,
and what kind of consequences we have after such \anoma breaking.
We defer the discussion on the former issue to future publication
and comment briefly on the latter here.

In general discrete symmetries survive
breaking of a continuous symmetry.
Suppose that the \anoma symmetry is broken
by a VEV of a scalar field with the $U(1)$ charge $Nq$.
If other fields which remain massless after this breaking
have charges quantized in units of $q$,
a discrete $Z_N$ subgroup of the original \anoma is left unbroken.
This $Z_N$ symmetry has $Z_N$ anomaly which comes from two sources.
As was discussed in Ref.~\cite{discrete:IR}
for breaking of an anomaly free $U(1)$,
a discrete gauge anomaly arises by integrating the matter fields
which gain mass terms through the symmetry breaking.
A new contribution arises from the original $U(1)$ anomaly.
With $\Tr'$ denoting a summation over massless fields 
after the \anoma breaking,
the $Z_N^3$ anomaly can be written as 
\be
\1{3}\Tr'Q_A^3=8\pi^2 \dGS 
+\1{3}\left(mN+\eta n{N^3\over 8}\right) \ ,
\label{discrete anomaly:cubic}
\ee
where $m$ and $n$ are some integers
and $\eta =1,\,0$ for $N=\,$even, odd, respectively.
The first term is the contribution from the $U(1)$ anomaly
and is proportional to the Green-Schwarz coefficient $\dGS$.
The second term is the contribution from massive fields,
for which we have used the formula given in Ref.~\cite{discrete:IR}.
Similarly the discrete version of the mixed gravitational anomaly
is calculated to be
\be
\1{24}\Tr'Q_A=8\pi^2 \dGS
+\1{24}\left(pN+\eta q\2{N}\right) \ ,
\label{discrete anomaly:grav}
\ee
where $p$ and $q$ are integers.
The integers $m$, $n$, $p$ and $q$ depend on the flat direction
along which the \anoma breaks.
If these $Z_N$ anomalies are to be cancelled
by the Green-Schwarz mechanism\cite{coping,discrete:Ib},
we should have a relation
\be
\1{3}\Tr'Q_A^3 = \1{24}\Tr' Q_A \qquad {\rm mod} \ N \ ,
\ee
which leads to the following constraint
\be
{8m-p \over 24}+\eta {2N^2-q \over 48}=\,{\rm integer} \ .
\ee
We can regard this as a constraint on possible flat directions.
Note that $\dGS$ cancels out and so
this constraint is independent of the original anomaly.
We can also calculate mixed $Z_N$ anomalies for gauge symmetries,
and have further constraints if these anomaly coefficients 
for $Z_N$ symmetry are to be universal.

Note that the \anoma breaking sometimes triggers
the further gauge symmetry breaking
if the scalar fields which develop the VEV's are 
charged under other gauge groups $G$.
This is the mechanism
which is widely used to reduce the rank of gauge group 
in semi-realistic string models\cite{degenerate,reduce,U1 charge}.
In such a case,
some linear combination of $U(1)$ and $G$ survives
and the discrete symmetry in the above discussion may be
taken to be orthogonal to it.

In addition to gauge symmetries,
there is another type of symmetries which are broken
in the course of the \anoma breaking.
We are interested in discrete $R$ symmetries and
let us consider them in $Z_3$ orbifold models as an example.
The $Z_3$ orbifold is symmetric 
under the independent $Z_3$ rotation of each complex plane as 
\ba
X^i&\longrightarrow&e^{2\pi i \wt{v}^i}X^i \ ,
\label{R:X}
\ea
where $\wt{v}^i=(n^1,n^2,n^3)/3$
with arbitrary integers $0\le n^i<3$.
The right-moving world-sheet supersymmetry requires that
the fermionic string coordinates $\psi^i\sim e^{iH^i}$
should be rotated simultaneously,
and these rotations are realized by the independent shift
of each $H^{i=1,2,3}$ as
\ba
H^i&\longrightarrow&H^i+2\pi\wt{v}^i \ ,
\label{R:H}
\ea
which rotates the space-time supercharge $Q\sim\prod_ie^{i H^i/2}$ as
\be
Q\ \longrightarrow\ 
e^{\pi i\sum_i \wt{v}^i}Q
\,\left(\,=e^{\3{\pi i}\sum_i n^i}Q\,\right) \ .
\label{R:supercharge}
\ee
Unless $\sum_i \wt{v}^i$ is an even integer,
these discrete symmetries (\ref{R:X}) and (\ref{R:H})
do not commute with the space-time supersymmetry
and are $R$ symmetries\cite{DS,naturalness}.

These $R$ symmetries are generated by
the rotation by $e^{2\pi i/3}$ of the $i$-th complex plane.
We call such generating element $R_i$.
The $R_i$ charge of a state can be read off as follows.
Massless scalar fields in the untwisted sector,
generally denoted by $U_{1,2,3}$,
have $SO(8)$ momenta $p^t=(p^i\,;\,0)$
with $p^i=(1,0,0)$, $(0,1,0)$ and $(0,0,1)$, respectively,
and the massless twisted field $T$ has $\wt{p}^t=(1,1,1\,;\,0)/3$.
These fields transform under the $R_i$ as
\be
R_i\,:\ 
U_j\ \longrightarrow\ e^{\3{2\pi i}\delta_{ij}}U_j \ , \qquad
T\ \longrightarrow\ e^{{2\pi i \over 9}}T \ ,
\ee
from which we see that each $R_i$ generates $Z_{18}$ symmetry.

The $R$ symmetries (\ref{R:X}) and (\ref{R:H}) can be accompanied
by discrete rotations of the left-moving gauge coordinates.
In the fermionic formulation, they are $Z_3$ rotations of
$\lambda^{\hat I}\sim e^{i\swb{X}^{\hat I}}$,
which are equivalently realized by discrete shifts of
the bosonic coordinates as
\ba
\wb{X}^{\hat I}&\longrightarrow&
\wb{X}^{\hat I}+2\pi\wt{V}^{\hat I} \ .
\label{R:G}
\ea
Here $\wt{V}^{\hat I}=n^{\hat I}/3$ with integers $0\le n^{\hat I}<3$
if the ${\hat I}$-th component of the shift $V^{\hat I}$ is non-vanishing.
[For vanishing component $V^{\hat I}=0$,
we should take $\wt{V}^{\hat I}=0$
since eq.~(\ref{R:G}) no longer give a symmtry transformation.]
In the (2,2) vacuum,
in order to preserve the left-moving world-sheet supersymmetry,
we should associate such $E_8\times E'_8$ rotation
that is realized by $\wt{V}^{\hat I}=(\wt{v}^i,0^5\,;\,0^8)$
with the discrete rotations of $X^i$ and $\psi^i$.
In (0,2) vacua, however,
the discrete rotations (\ref{R:G}) are independent and
we are free to associate them with eqs.~(\ref{R:X}) and (\ref{R:H}).
Note that the discrete symmetries (\ref{R:G}) themselves are
nothing but discrete parts of gauge symmetries.

These $R$-symmetries (\ref{R:H}) and discrete parts (\ref{R:G})
of gauge symmetries are generally broken
when the \anoma breaks along a flat direction.
Suppose that the matter field
which develops the VEV to cancel the Fayet-Iliopoulos term
has $SO(8)$ and $E_8\times E_8'$ momenta
$\wt{p}^i$ and $\wt{P}^{\hat I}$, respectively.
The corresponding vertex operator transforms
under eqs.~(\ref{R:X}), (\ref{R:H}) and (\ref{R:G}) as\footnote{
{}For an oscillated state,
the phase $\exp(2\pi iN_{\L})$ should be multiplied.
Note also that
if we are to transform the twist field at the same time,
we have the corresponding phase from the twisted ground state.
In such a case,
the $R$ charge of the state can be read off
by the formula similar to the GSO phase (\ref{GSO:theta}).
}
\be
V_{\R}V_{\L} \ \longrightarrow\ 
e^{2\pi i\left(\wt{P}^{\hat I}\wt{V}^{\hat I}
-\wt{p}^{i}\wt{v}^{i}\right)} V_{\R}V_{\L} \  .
\label{Rcharge}
\ee
Thus
the original $R$ symmetries generated by $R_i$ are generally broken,
and the surviving $R$ symmetries are such combinations of $R_i$
and discrete symmetries that satisfy
\be
\wt{P}^{\hat I}\wt{V}^{\hat I}-\wt{p}^{i}\wt{v}^{i}
=\,{\rm integer} \ .
\label{survive}
\ee
If several matter fields develop their VEV's,
the shifts $\wt{v}^i$ and $\wt{V}^{\hat I}$ of
the surviving $R$ symmetry should satisfy this relation
for each pair of $\wt{p}^i$ and $\wt{P}^{\hat I}$.

Let us illustrate the situation
by taking the $Z_3$ orbifold model with No.~4 shift,
whose flat directions were analyzed in Ref.~\cite{degenerate}.
This model has the gauge group $SU(9)\times SO(14)' \times U(1)'$
and the massless matter content is shown in Appendix A.
{}First consider the simplest flat direction where
the negatively charged field $(1,14')_{-1}$ in the untwisted sector
develops the VEV which cancels the positive Fayet-Iliopoulos term
and breaks $SO(14)'$ simultaneously.
In this case,
the discrete gauge symmetry which survives the breaking
is $Z_6$ under which
\be
\phi_{64'}\ \longrightarrow\ e^{k\pi i}\phi_{64'} \ , \qquad
\phi_{9}\ \longrightarrow\ e^{\3{4\pi i}k}\phi_{9} \ ,
\ee
where $k=0,\,\cdots,\,5$.
On the other hand,
if the field $(1,14')$ which develops the VEV has $p^i=(1,0,0)$,
the $R$ symmetries generated by $R_1$ are broken
while the $R_{2,3}$ are left unbroken.
Although the $R_1$ is broken,
we can find a new unbroken $R$-symmetry 
by combining the discrete part of gauge symmetries so as to
satisfy eq.~(\ref{survive}),
\ie, by combining the symmetry (\ref{R:G}) generated by
$\wt{V}^{\hat I}=(0^8\,;\,2/3,0^7)$.
Notice that this is exactly the basis for the \anom,
and thus the surviving $R$ symmetry is a combination of
the broken $R_1$ and the broken part of \anoma gauge symmetry.

This $Z_3$ orbifold model has another flat direction,
where $(9,1')_{2/3}$ as well as $(1,14')_{-1}$ develop their VEV's.
These $(9,1')_{2/3}$ fields come from the twisted sector
and have $\wt{p}^i=(1,1,1)/3$.
Switching on their VEV's breaks the above new $R$-symmetry
associated with the \anom.
Even in this case, however, the unbroken $R$-symmetry
can be found by further combining the center of the $SU(9)$
generated by $\wt{V}^{\hat I}=(1,1,1,1,2,0,0,0\,;\,0^8)/3$.

The other $Z_3$ orbifold models also have unbroken $R$-symmetries 
after symmetry breaking along flat directions.
This analysis can be extended to $Z_3$ orbifold models
with Wilson lines and other orbifold models.
Unbroken $R$-symmetries, in general, contain
discrete subgroups of broken gauge symmetries
including the \anom.

\section{Conclusion and discussion}

We have studied the origin of \anoma gauge symmetry
in the orbifold construction of four-dimensional string models.
By utilizing the universal nature of the anomaly in string theory,
we have derived several conditions for the absence of $U(1)$ anomaly.
We also have found several discrete symmetries
which guarantee the cancellation of anomaly.
We have then presented a procedure for classifying
the orbifold string models which possess an \anoma
and for identifying the true basis of the \anom.

We have found several constraints on the \anom,
which may be regarded as the first step
to have the complete criteria for the appearance of an \anom.
These constraints are rather independent
of the detailed structure of models
and therefore enable us to conclude the absence of an \anoma
before going into the analysis of massless spectra.
According to them,
one can conclude the absence of an \anoma by the following reasons:
(i) the absence of the visible-hidden sector mixing
in the twisted sectors (as in the models with a large gauge group),
(ii) the existence of discrete symmetries, which can be found
by examining the relation between the shift and Wilson lines
(such as the orthogonality).
The former is powerful enough to allow us
to classify the models in the absence of a Wilson line.
Such a classification is the basis for
more general and detailed analysis in the presence of Wilson lines.
On the other hand, the latter plays an important role
complimentary to such a detailed analysis.

One of the main results of the present paper is to give
a general procedure for classifying and identifying 
the \anoma in orbifold string models.
According to our procedure, the problem is reduced to
the classification of models in the absence of a Wilson line.
Once we work out this restricted class of models and identify
what type of shifts and twisted sectors lead to an \anom,
we can easily extend the analysis to the case with Wilson lines.
This greatly simplifies the actual analysis of an \anoma
since we can avoid the tedious calculation of $U(1)$ charges.

Our procedure is based on the fact that
an orbifold model is constructed by modular invariant combinations
of the twisted subsectors corresponding to fixed points
and that any such subsector in the presence of Wilson lines
is equivalent by $E_8\times E_8'$ automorphisms
to some twisted sector in the absence of a Wilson line.
The actual analysis is quite easy when such an equivalence
is realized by a trivial $E_8\times E_8'$ automorphism
as we demonstrated in concrete examples.
When nontrivial $E_8\times E_8'$ automorphisms are needed,
the analysis will be somehow involved.
The investigation on this point will be given elsewhere.

The fact that
an orbifold model in a Wilson line background is constructed
by assigning various types of shifts to twisted subsectors
and by combining such subsectors in a modular invariant way,
might be interesting by itself.
We have observed the similarity to the construction of a fiber bundle.
More importantly, this fact makes it clear that
the origin of an \anoma can be traced back to
the orbifold twist itself which is realized by the shift $V^{\hat I}$
in the Abelian embedding adopted here.
This may not be surprising since it is this twist operation
that gives rise to the chiral structure of the models.
[In this respect,
it may be interesting to recall that
$E_8\times E_8$ heterotic string theory itself may be regarded
as an orbifold in the M-theoretical picture\cite{horava-witten}.
It might also be interesting to speculate that
any chiral structure could be traced to an orbifold in some sense.]

Several problems are left untouched in the present paper.
One of the most important problems is
to give a general characterization of
the flat directions along which the \anoma breaks.
Such an analysis of the flat directions are indispensable
for many applications of the \anoma to the realistic model building,
such as the construction of string models with
realistic gauge groups and matter contents,
fermion mass matrices\cite{texture:IR,texture:others,texture2},
supersymmetry breaking\cite{SUSY-breaking:mech,SUSY-breaking:model},
and the calculation of soft breaking terms\cite{anoma,KK,mass:others}.
Our results, the classification of an \anoma in particular,
will be useful for such an analysis of the flat directions.
Also helpful will be the results of Ref.~\cite{flat} where
generic flat directions of $Z_{2n}$ orbifold models were worked out
(although the breaking of the \anoma was not taken into account).

In attempts to construct a realistic string model,
discrete symmetries including $R$ symmetries are important,
for instance, to constrain phenomenologically dangerous couplings.
We have discussed discrete symmetries
that survive the \anoma breaking and observed a possible relation
to the broken \anom.
A further study on such a remnant of the \anoma will be desired.

It is a remarkable possibility that the \anoma gauge symmetry
plays an important role in constructing a promising model of
supersymmetry breaking\cite{SUSY-breaking:mech,SUSY-breaking:model}.
If this is the case,
it is very important to search for a simple and concrete example
of the string model in which the supersymmetry breaking mechanism
of ref.~\cite{SUSY-breaking:mech} is realized.
{}For this purpose, we note that
there are many such orbifold models without a Wilson line
that possess an \anoma
and that these models deserve further study
even if they are not necessarily realistic by themselves.

Recently much work has been devoted
to understand nonperturbative aspects 
in supersymmetric gauge theories and string theories.
In particular, it was pointed out in Ref.~\cite{moduli} that
the nonperturbative effects of the form $e^{-aS}$
are constrained by the \anoma as well as discrete symmetries.
[This may be understood by observing that
if we use the field $e^{-S}$, the \anoma is linearly realized and
the Green-Schwarz coefficient $\dGS$ is just its \anoma charge.]
In this sense, the \anoma gauge symmetry might be related to
some of nonperturbative and universal aspects of the theory.
It is therefore important to study the \anoma
from the viewpoint of string duality and M-theory.

Our approach here to the \anoma can be extended to
other constructions of four-dimensional string models
such as Calabi-Yau compactifications\cite{CY}
and fermionic constructions\cite{fermi}.
In particular,
the absence of the visible-hidden sector mixing will provide
one of the criteria for anomaly free models in any construction.
On the other hand,
a new clue might be obtained by studying orbifold models
with the non-Abelian embedding.
The fact that the rank of gauge groups can be lowered
in such a construction\cite{nonabelian} might be related to
the gauge symmetry breaking triggered by the \anoma breaking.

\section*{Acknowledgements}
The authors are grateful to Y.~Kawamura and J.~Louis 
for useful discussions.
They would like to express their sincere thanks to
M.~Bando and I.A.~Sanda, the organizers of Ontake Summer Institute
where a part of the work was done.
H.~N. is grateful to the colleagues at Niigata University
for encouragement and to INS for hospitality.
H.~N. is supported in part
by the Grant-in-Aid for Scientific Research (\#~08740198)
from the Ministry of Education, Science and Culture of Japan.

\newpage

\appendix
\section{Examples of ``Anomalous'' $U(1)$}

Here we give two examples for the models with the \anom.
They are $Z_3$ orbifold models with No.~3 and 4 type shifts
and without Wilson lines.
As we argued in section~4, \anoma in all the $Z_3$ orbifold models
can essentially be traced to these two models
even after Wilson lines are added.

The model with No.~4 type shift $V^{\hat{I}}=(1^4,2,0^3)(2,0^7)'/3$
has the gauge group $SU(9)\times SO(14)'\times U(1)'_A$.
$U(1)'_A$ charges are calculated by the formula (\ref{charge:def})
as $Q'_A=V^{\hat I}_A\tilde P^{\hat I}$, 
where the $U(1)$ basis is given by $V_A=(0^8)(1,0^7)'$.
Note that with this choice of basis,
the level of $U(1)'_A$ is given by the formula (\ref{level:def})
as $k_A=2$.
The massless matter fields are given in an obvious notation by
\ba
\hbox{$U$-sector}&:&~~~~\,
3\times\left[\,(84,1')_{0}+(1,14')_{-1}
+\left(1,64_s'\right)_{\2{1}}\,\right] \ ,
\nonumber\\
\hbox{$T_1$-sector}&:&~~~
27\times\left[\,(9,1')_{\3{2}}\,\right] \ .
\label{matter:No.4}
\ea
Observe that
the untwisted matter $(84,1')$ has vanishing $U(1)'_A$ charge
while the twisted matter $(9,1')$ is $U(1)'_A$-charged,
signaling that $U(1)'_A$ is anomalous.
It is straightforward to check that 
we have
\be
\mathop{\Tr}_{~SU(9)}T(R)Q'_A = \mathop{\Tr}_{SO(14)'}T(R)Q'_A
=\1{6}\Tr Q'^3_A=\1{24}\Tr Q'_A=9 \ .
\label{GS:No.4}
\ee
With the rescaling $Q_A'\rightarrow\sqrt{2}Q_A'$,
the universal GS relation (\ref{universal}) is indeed satisfied
with $8\pi^2\sqrt{2}\dGS=9$.
As discussed in section~3, this universality,
which guarantees the anomaly cancellation by GS mechanism,
enables us to search for the \anom.

The second example is the model with No.~3 type shift:
$V^{\hat{I}}=(1^2,0^6)(2,0^7)'/3$.
The gauge group is $E_7\times U(1)_A\times SO(14)'\times U(1)_B'$
and $U(1)$ charges are
\ba
Q_A=\sum_{\hat{I}}V_A^{\hat{I}}\wt{P}^{\hat{I}}&,& \qquad
V_A^{\hat{I}}=\left(1,1,0^6\right)\left(0,0^7\right)' \ ,
\nonumber\\
Q_B'=\sum_{\hat{I}}V_B^{\hat{I}}\wt{P}^{\hat{I}}&,& \qquad
V_B^{\hat{I}}=\left(0,0,0^6\right)\left(1,0^7\right)' \ .
\label{UAB}
\ea
The massless matter fields with charge $(Q_A,Q_B')$ are
\ba
\hbox{$U$-sector}&:&~~~~\,
3\times\left[\,
(56,1')_{1,0}+(1,1')_{-2,0}+(1,14')_{0,-1}+(1,64_s')_{0,\2{1}}
\,\right] \ ,
\nonumber\\
\hbox{$T_1$-sector}&:&~~~
27\times\left[\,(1,14')_{\3{2},-\3{1}}+(1,1')_{-\3{4},\3{2}}
+3\times (1,1')_{\3{2},\3{2}}\,\right] \ ,
\label{matter:No.3}
\ea
where the last singlets with the degeneracy factor $27\times 3$
come from oscillated states $N_{\L}=1/3$.
As argued in section~3,
the hidden $U(1)_B'$ is not anomalous since
no $\ul{56}$ appears in the twisted sector and
there is no mixing between $E_7$ and $U(1)_B'$.
On the other hand,
the presence of $\ul{14}'$ in the twisted sector signals that
the visible $U(1)_A$ is anomalous.
Actually we have
\be
\mathop{\Tr}_{~E_7}T(R)Q_A = \mathop{\Tr}_{SO(14)'}T(R)Q_A
=\1{2}\Tr Q_B'^2Q_A=\1{12}\Tr Q_A^3=\1{24}\Tr Q_A=18 \ .
\label{GS:No.3}
\ee
Note that in the $U(1)$ basis (\ref{UAB}),
the levels of $U(1)_A$ and $U(1)_B'$ are $k_A=4$ and $k_B=2$,
respectively. 
Then after the rescaling
$Q_A\rightarrow 2Q_A$ and $Q_B'\rightarrow\sqrt{2}Q_B'$,
we find the universal GS relation of the form (\ref{universal})
with $\dGS=9/8\pi^2$.

\section{Classification of $Z_4$ orbifold models}

Here we classify $Z_4$ orbifold models with \anom. 
Ten independent $Z_4$ shifts are shown in Table~\ref{table:Z4 shift}
including the trivial one\cite{shift}.
Modular invariant pairs of shifts $(V^I\,;\,V^{I'})$
of $Z_4$ orbifold models are classified\cite{Z4}
into twelve independent pairs as shown in Table~\ref{table:Z4 mixing},
whose third column shows the allowed combinations of
$E_8$ shift $V^I$ and $E'_8$ shift $V^{I'}$
in terms of the corresponding numbers
of the first column in Table ~\ref{table:Z4 shift}.
The fourth and fifth columns of Table~\ref{table:Z4 mixing} show
such massless matter fields in $T_1$ and $T_2$ twisted sectors
that give rise to visible-hidden sector mixing and contribute to
the anomaly with respect to the largest gauge group.
All of these fields correspond to non-oscillated states.

The models No.~8, 9 and 10 have no $U(1)$ gauge group.
The models No.~1 and 2 have $E'_8$ and $E'_7 \times SU(2)'$,
respectively as the gauge group in the hidden sector
and thus are anomaly free
owing to the reason discussed in subsection~3.1.
The remaining seven models have \anom.
There appear several new features which are absent
in the case of $Z_3$ orbifold models discussed in section~3.
{}Firstly,
the $T_2$ sector has a conjugate pair of massless matter fields,
\ie, $R$ and $\wb{R}$, but these fields
generally have different degeneracies\rlap.\footnote{
See Ref.~\cite{flat}, where generic massless spectra
in the $T_n$ sector of $Z_{2n}$ orbifold models are classified.}
Secondly, as we describe shortly,
a linear combination of the visible $U(1)$ and hidden $U(1)'$ 
corresponds to a true basis of the \anoma in the model No.~5.
On the other hand, only the $U(1)'$ group is anomalous in the model 
No.~4 as well as No.~11.
Thirdly in the model No.~7,
the twisted sector $T_1$ has no contribution to the anomaly
despite the fact that the whole model has the \anom.
After a Wilson line is included, therefore,
a subsector of the $T_1$ sector does not contribute to an anomaly
even if it has the No.~7 type shift as the equivalent shift.
Similarly the $T_2$ sector has no contribution to the anomaly
in the models No.~3, 11 and 12.
This is the reason why
we need to work out which twisted sector contributes to the anomaly
before including a Wilson line.

As an example, we explicitly give the result on the model No.~5,
whose gauge group is $SO(12)\times SU(2)\times U(1)_\alpha%
\times SO(14)'\times U(1)'_\beta$.
The levels of $U(1)_\alpha$ and $U(1)'_\beta$
are $k_\alpha=4$ and $k_\beta=8$
if we use the basis given in Table~\ref{table:Z4 shift}.
The massless matter content is given by
\ba
\hbox{$U_1$}:&&\!\!
2\times\left[\,
(12,2;1')_{-1,0}+(32,1;1')_{1,0}+(1,1;64'_s)_{0,2}
     \,\right] \ ,
\nonumber\\
\hbox{$U_2$}:&&\!\!
1\times\left[\,
(32,2;1')_{0,0}+(1,1;1')_{\pm 2,0}+(1,1;14')_{0,\pm 2}
     \,\right] \ ,
\nonumber\\
\hbox{$T_1$}:&&\!\!\!\!\!
16\times\left[\,
(12,1;1')_{-\1{2},1}+2\times (1,2;1')_{\1{2},1}
      \,\right] \ ,
\nonumber\\
\hbox{$T_2$}:&&\!\!
4\times
\left[\,(1,1;14')_{-1,0}+(1,1;1')_{1,\pm 2}\,\right]
\nonumber\\
&&\!\!\!\!\!\!\!
+6\times
\left[\,(1,1;14')_{\mp 1,0}+(1,1;1')_{\pm(1,\pm 2)}\,\right]\ ,
\label{matter:Z4:No.5}
\ea
where the degeneracy factors are determined
by the generalized GSO projection.
Notice that
all fields in the $U_2$ sector and six copies of the fields
in the $T_2$ sector form $N=2$ hypermultiplets
and therefore do not contribute to anomaly.
A straightforward calculation shows that
\ba
&&\mathop{\Tr}_{~G_a}T(R)Q_{\alpha}=\1{8}\Tr Q'^2_{\beta}Q_{\alpha}
=\1{12}\Tr Q^3_{\alpha}=\1{24}\Tr Q_{\alpha}=-4 \ ,
\nonumber\\
&&\mathop{\Tr}_{~G_a}T(R)Q'_{\beta}=\1{4}\Tr Q^2_{\alpha}Q'_{\beta}
=\1{24}\Tr Q'^3_{\beta}=\1{24}\Tr Q'_{\beta}=16 \ ,
\label{GS:Z4:No.5}
\ea
where $G_a$ stands for $SO(12)$, $SU(2)$ or $SO(14)'$.
We have explicitly confirmed the claim that
the universal GS relation does not depend on
the choice of $U(1)$ basis.
The true basis of the anomalous $U(1)_A$ is then found to be
$Q_A\propto -Q_{\alpha}+2Q'_{\beta}$,
which is orthogonal to the anomaly free combination
$Q_B\propto 4Q_{\alpha}+Q'_{\beta}$.

In $Z_4$ orbifold models with Wilson lines,
each subsector corresponding to the fixed point has the total shift
of the form $(kV^I+m^ia^I_i\,;\,kV^{I'}+m^ia^{I'}_i)$,
which are also classified into the above twelve types
up to $E_8 \times E'_8$ automorphisms.
As was commented in section~4, one subsector can have
a different type of the equivalent shift from others.
Thus the anomaly in each subsector is completely classified
by using the results given here in Table~\ref{table:Z4 mixing},
and the true \anoma basis of the whole model
can be obtained as a linear combination of 
the would-be \anoma bases of several subsectors.
We can extend these study to other orbifold models.

\begin{table}
\begin{center}
\begin{tabular}
{@{\vrule width 1pt \ }c|c|c|c@{\ \vrule width 1pt}}
\noalign{\hrule height 1pt}
\#         & Shift ($4V^I$) & Gauge group & $U(1)$ basis
\\ 
\noalign{\hrule height 1pt}
\hline \hline
\noalign{\hrule height 1pt}%
0  	& $(00000000)$ & $E_8$ & \\ \hline
1  	& $(22000000)$ & $E_7\cdot SU_2$ &  \\ \hline
2  	& $(11000000)$ & $E_7\cdot U_1$ 
	& $(11000000)$ \\ \hline
3  	& $(21100000)$ & $E_6\cdot SU_2\cdot U_1$ 
	& $(21100000)$ \\ \hline
4  	& $(40000000)$ & $SO_{16}$ &  \\ \hline
5  	& $(20000000)$ & $SO_{14}\cdot U_1 $ 
	& $(20000000)$ \\ \hline
6  	& $(31000000)$ & $SO_{12}\cdot SU_2\cdot U_1$
	& $(1\!-\!\!1000000)$ \\ \hline
7  	& $(22200000)$ & $SO_{10}\cdot SU_4$ &  \\ \hline
8  	& $(31111100)$ & $SU_8\cdot SU_2$ &  \\ \hline
9  	& $(1111111\!-\!\!1)$ & $SU_8\cdot U_1$ 
	& $(1111111\!-\!\!1)$ \\
\noalign{\hrule height 1pt}
\end{tabular}
\end{center}
\caption{Shifts for $Z_4$ orbifold models}
\label{table:Z4 shift}
\end{table}

\begin{table}
\begin{center}
\begin{tabular}
{@{\vrule width 1pt \ }c|c|c|c|c@{\ \vrule width 1pt}}
\noalign{\hrule height 1pt}
No. & Gauge group & $(V^I;V^{I'})$ & $T_1$ & $T_2$ 
\\ 
\noalign{\hrule height 1pt}
\hline \hline 
\noalign{\hrule height 1pt}
1 	& $E_6\cdot SU_2\cdot U_1\cdot E'_8$ 
	& $\model{3}{0}$ 
	&
	& 
\\ \hline
2 	& $E_6\cdot SU_2\cdot U_1\cdot E'_7\cdot SU'_2$ 
	& $\model{3}{1}$
	&
	& 
\\ \hline
3 	& $SO_{16}\cdot E'_6\cdot SU'_2\cdot U'_1$ 
	& $\model{4}{3}$
	& $(16;1',1')_{3/2}$
	& 
\\ \hline
4 	& $SO_{14}\cdot U_1\cdot E'_7\cdot U'_1$ 
	& $\model{5}{2}$
	& $(14;1')_{-1,1/2}$
	& $(14;1')_{0,\pm 1}$
\\ \hline
5	& $SO_{12}\cdot SU_2\cdot U_1\cdot SO'_{14}\cdot U'_1$
	& $\model{6}{5}$
	& $(12,1;1')_{-1/2,1}$
	& $(1,1;14')_{\pm 1,0}$
\\ \hline
6 	& $SO_{10}\cdot SU_4\cdot E'_7\cdot U'_1$ 
	& $\model{7}{2}$
	& $(16,1;1')_{1/2}$ 
	& $(10,1;1')_{\pm 1}$ 
\\ \hline
7 	& $SO_{10}\cdot SU_4\cdot SO'_{12}\cdot SU'_2\cdot U'_1$
	& $\model{7}{6}$ 
	& 
	& $(10,1;1',1')_{\pm 1}$ 
\\ \hline
8 	& $SU_8\cdot SU_2\cdot E'_8$ 
	& $\model{8}{0}$
	&
	&  
\\ \hline
9 	& $SU_8\cdot SU_2\cdot E'_7\cdot SU'_2$
	& $\model{8}{1}$
	&
	& 
\\ \hline
10 	& $SU_8\cdot SU_2\cdot SO'_{16}$ 
	& $\model{8}{4}$
	&
	& 
\\ \hline
11 	& $SU_8\cdot U_1\cdot E'_6\cdot SU'_2\cdot U'_1$ 
	& $\model{9}{3}$
	& $(8;1',1')_{-1,3/2}$
	& 
\\ \hline
12 	& $SU_8 \cdot U_1\cdot SU'_8\cdot SU'_2$ 
	& $\model{9}{8}$
	& $(1;8',1')_2$
	& 
\\ 
\noalign{\hrule height 1pt}
\end{tabular}
\end{center}
\caption{Visible-hidden sector mixing in $Z_4$ orbifold models}
\label{table:Z4 mixing}
\end{table}


\newpage


\begin{thebibliography}{99}
\bibitem{GS}
        M.~B.~Green and J.~H.~Schwarz, \PL{149B}{117}{84}.
\bibitem{moduli}
	T.~Banks and M.~Dine, \PR{D53}{5790}{96}.
\bibitem{FI}
         P.~Fayet and J.~Iliopoulos, \PL{51B}{461}{74};
\\
        W.~Fischler, H.~P.~Nilles, J.~Polchinski, S.~Raby
        and L.~Susskind, \PRL{47}{757}{81}.
\bibitem{DSW}
	E.~Witten, \PL{149B}{351}{84};
\\
        M.~Dine, N.~Seiberg and E.~Witten, \NP{B289}{589}{87};
\\
        W.~Lerche, B.~E.~W.~Nilsson and A.~N.~Schellekens,
        \NP{B289}{609}{87}.
\bibitem{DSWii}
        J.~Atick, L.~Dixon and A.~Sen, \NP{B292}{109}{87};
\\
        M.~Dine, I.~Ichinose and N.~Seiberg, \NP{B293}{253}{87}.
\bibitem{degenerate}
        A.~Font, L.E.~Ib\'a\~nez, H.P.~Nilles and F.~Quevedo,
        \NP{B307}{109}{88}; \NP{B310}{764}{88}.
\bibitem{reduce}
        A.~Font, L.E.~Ib\'a\~nez, H.P.~Nilles and F.~Quevedo, 
        \PL{B210}{101}{88}; \PL{B213}{564}{88};
\\
        J.A.~Casas and C.~Mu\~noz,
        \PL{209B}{214}{88}; \PL{214B}{63}{88}.
\bibitem{U1 charge}
        J.A.~Casas, E.K.~Katehou and C.~Mu\~noz,
        \NP{B317}{171}{89}.
\bibitem{IB}
        L.~E.~Ib\'a\~nez, \PL{B303}{55}{93}.
\bibitem{texture:non-anomalous}
        M.~Leurer, Y.~Nir and N.~Seiberg,
        \NP{B398}{319}{93}; \NP{B420}{468}{94};
\\
        Y.~Nir and N.~Seiberg, \PL{B309}{337}{93}.
\bibitem{texture:IR}
        L.E.~Ib\'a\~nez and G.G.~Ross, \PL{B332}{100}{94}.
\bibitem{texture:others} 
	V.~Jain and R.~Shrock, \PL{B352}{83}{95};
\\
        P.~Bin\'etruy and P.~Ramond, \PL{B350}{49}{95};
\\
        E.~Dudas, S.~Pokorski and C.A.~Savoy, \PL{B356}{45}{95}.
\bibitem{texture2}
        T.~Kobayashi, \PL{B354}{264}{95}; \PL{B358}{253}{95};
\\
	T.~Kobayashi and Z.Z.~Xing, INS-Rep-1162, hep-ph/9609486.
\bibitem{Di}
        M.~Drees, \PL{181B}{279}{86};
\\
        J.~S.~Hagelin and S.~Kelley, \NP{B342}{95}{90};
\\
        Y.~Kawamura, H.~Murayama and M.~Yamaguchi,
        \PL{B324}{94}{94}.
\bibitem{KMY}
        Y.~Kawamura, H.~Murayama and M.~Yamaguchi,
        \PR{D51}{1337}{95}.
\bibitem{anoma}
        H.~Nakano, Kyoto preprint KUNS-1257, HE(TH) 94/05,
        hep-th/9404033;
\\
	See also, H.~Nakano, \PTPS{\bf 123}{387}{96}.
\bibitem{KK}
        Y.~Kawamura and T.~Kobayashi, \PL{B375}{141}{96};
	INS-Rep-1153, hep-ph/9608233;
\\
	Y.~Kawamura, T.~Kobayashi and T.~Komatsu,
	INS-Rep-1161, hep-ph/9609462.
\bibitem{mass:others}
	E.~Dudas, S.~Pokorski and C.A.~Savoy, \PL{B369}{255}{96};
\\
	E.~Dudas, C.~Grojean, S.~Pokorski and C.A.~Savoy,
	Saclay T96/065, hep-ph/9606383.
\bibitem{SUSY-breaking:mech}
	G.~Dvali and A.~Pomarol, \PRL{77}{3728}{96};
\\
	P.~Binetruy and E.~Dudas, LPTHE-Orsay 96/60, hep-th/9607172.
\bibitem{SUSY-breaking:model}
	R.N.~Mohapatra and A.~Riotto, hep-ph/9608441.
\bibitem{cosmology}
        J.A.~Casas, J.M.~Moreno, C.~Mu\~noz and M.~Quiros,
        \NP{B328}{272}{89}.
\bibitem{inflation}
	P.~Binetruy and G.~Dvali, CERN-TH/96-149, hep-ph/9606342;
\\
	E.~Halyo, SU-ITP-96-28, hep-ph/9606432.
\bibitem{splitting}
	G.~Dvali and S.~Pokorski, CERN-TH/96-286, hep-ph/9610431.
\bibitem{DS}
        M.~Dine and N.~Seiberg, \NP{B306}{137}{88}.
\bibitem{naturalness}
	A.~Font, L.E.~Ib\'a\~nez, H.P.~Nilles and F.~Quevedo, 
        \PL{B213}{274}{88}.
\bibitem{DSWW}
	X.-G.~Wen and E.~Witten, \PL{166B}{397}{86};
\\
        M.~Dine, N.~Seiberg, X.-G.~Wen and E.~Witten,
	\NP{B278}{789}{86}; \NP{B289}{319}{87}.
\bibitem{Distler-Greene}
	J.~Distler and B.~Greene, \NP{B304}{1}{88}.
\bibitem{Distler-Kachru}
	J.~Distler and S.~Kachru, \NP{B413}{213}{94}.
\bibitem{Silverstein-Witten}
	E.~Silverstein and E.~Witten, \NP{B444}{161}{95}.
\bibitem{orbifold:original}
        L.~Dixon, J.~Harvey, C.~Vafa and E.~Witten,
        \NP{B261}{651}{85}; \NP{B274}{285}{86}.
\bibitem{Orbi1}
        L.E.~Ib\'a\~nez, J.~Mass, H.P.~Nilles and F.~Quevedo,
        \NP{B301}{157}{88}.
\bibitem{Orbi2}
        Y.~Katsuki, Y.~Kawamura, T.~Kobayashi, N.~Ohtsubo, Y.~Ono
        and K.~Tanioka, \NP{B341}{611}{90}.
\bibitem{Orbi3}
        T.~Kobayashi and N.~Ohtsubo, \IJMP{A9}{87}{94}.
\bibitem{shift}
        Y.~Katsuki, Y.~Kawamura, T.~Kobayashi, N.~Ohtsubo, 
        and K.~Tanioka,
        \PTP{82}{171}{89}.
\bibitem{orbifold-CFT}
        L.~Dixon, D.~Friedan, E.~Martinec and S.~Shenker,
        \NP{B282}{13}{87}.
\bibitem{KO:linear comb}
        T.~Kobayashi and N.~Ohtsubo, \PL{B245}{441}{90}.
\bibitem{realistic}
        A.~Font, L.E.~Ib\'a\~nez, F.~Quevedo and A.~Sierra,
	\NP{B331}{421}{90}.
\bibitem{GSO}
        I.~Senda and A.~Sugamoto, \NP{B302}{291}{88}.
\bibitem{KO:WL}
        T.~Kobayashi and N.~Ohtsubo, \PL{B257}{56}{91}.
\bibitem{FI:R}
	B.~Barbieri, S.~Ferrara, D.V.~Nanopoulos and K.S.~Stelle,
	\PL{113B}{219}{82};
\\
	K.S.~Stelle and P.C.~West, \NP{B145}{175}{78};
\\
	S.~Ferrara, L.~Girardello, T.~Kugo and A.~Van~Proeyen,
	\NP{B223}{191}{83}.
\bibitem{U1 mixing}
	B.~Holdom, \PL{B166}{196}{86};
\\
	K.R.~Dienes, C.~Kolda and J.~March-Russel,
	IASSNS-HEP-96/100, hep-ph/9610479.
\bibitem{SW:anomaly}
        A.N.~Schellekens and N.P.~Warner,
        \NP{B287}{317}{87}.
\bibitem{reduction}
        J.A.~Casas, M.~Mondragon and C.~Mu\~noz,
	\PL{B230}{63}{89}.
\bibitem{CP}
	C.S.~Lim, \PL{B256}{233}{91};
\\
	M.~Dine, R.G.~Leigh and D.A.~MacIntire,
	\PRL{69}{2030}{92};
\\
	K.~Choi, D.B.~Kaplan and A.E.~Nelson,
	\NP{B391}{515}{93};
\\
	T.~Kobayashi and C.S.~Lim,
	\PL{B343}{122}{95}.
\bibitem{discrete:IR}
       L.E.~Ib\'a\~nez and G.G.~Ross, \PL{B260}{291}{91}.
\bibitem{coping}
	T.~Banks and M.~Dine, \PR{D45}{1424}{92}.
\bibitem{discrete:Ib}
       L.E.~Ib\'a\~nez, \NP{B398}{301}{93}.
\bibitem{horava-witten}
	P.~Ho\v rava and E.~Witten,
	\NP{B460}{506}{96}; hep-th/9603142.
\bibitem{flat}
        Y.~Kawamura and T.~Kobayashi, preprint, INS-Rep-1149, 
        hep-th/9606189,
	to be published in {\it Nucl}.~{\it Phys}.~{\bf B}.
\bibitem{CY}
	P.~Candelas, G.~Horowitz, A.~Strominger and E.~Witten, 
	\NP{B258}{46}{85}.
\bibitem{fermi}
	H.~Kawai, D.C.~Lewellen and A.N.~Shellekens, 
	\PRL{57}{1853}{86}; \NP{B288}{1}{87};
\\
	I.~Antoniadis, C.~Bachas and C.~Kounnas, 
	\NP{B289}{87}{87}.
\bibitem{nonabelian}
        L.E.~Ib\'a\~nez, H.P.~Nilles and F.~Quevedo,
	\PL{B192}{332}{87}.
\bibitem{Z4}
        Y.~Katsuki, Y.~Kawamura, T.~Kobayashi, N.~Ohtsubo, Y.~Ono
        and K.~Tanioka, \PL{B218}{169}{89}.
\end{thebibliography}
\end{document}